\documentclass[english,12pt]{article}
\usepackage[cp1251]{inputenc}
\usepackage{babel}
\usepackage{amsmath, amssymb}
\usepackage{axodraw}
\usepackage{epsfig}
\begin{document}
\title{
Investigation of some processes of multiple lepton and boson birth on linear colliders of polarized photons.
}
\author{T. Shishkina\thanks{BSU, Minsk}}
\date{}
\maketitle
\begin{abstract}

The main possibilities of investigation of
leptons and bosons birth
in interaction of polarized photons are considered.
The usage of
$\gamma \gamma \to f \bar{f} [+\gamma]$ reactions for the luminosity measurement
on linear photon collider is analyzed.
The achievable precision of the luminosity measuring is considered and calculated.
The first-order QED correction to $\gamma\gamma\to l \bar{l}$
scattering is analyzed.
Differential cross section of process $\gamma\gamma\to 4l$ is
calculated using the helicity amplitudes method as well as
covariant method of precision calculations. All possible polarization
states of interacting particles are investigated under different cuts
of TESLA kinematics.
For the detection of deviations from SM predictions
at linear $\gamma\gamma$ colliders with centre of mass energies running to $1\, TeV$
the influence of three possible
anomalous couplings on the cross sections of $W^+W^-$
productions has been investigated.
The significant discrimination between various anomalous contributions is discovered.
The main contribution of high order electroweak effects is considered.
\end{abstract}

\mathchardef\vm="117
\mathchardef\um="11D
\mathchardef\E="245
\mathchardef\Mom="250
\def\z1{z{}'}
\def\t1{t{}'}
\def\u1{u{}'}
\def\s1{s{}'}
\def\m2{m^2}
\def\d12{\frac{1}{2}}
\def\lfr#1#2{\ln{\frac{#1}{#2}}}
\def\bracket#1{\left({#1}\right)}
\def\bra#1{\bracket{#1}}
\def\spr#1#2{\, #1 \! \cdot \! #2 \,}
\def\ddx#1{\frac{1}{#1}}

\section{Introduction}

The Standard Model (SM) is able to describe all
experimental data up to now
with typical precision around
one per mille. Nevertheless the Model is
not the final theory valid up
to very high scales and
at linear collider that can run at centre of mass energies around 1 TeV
one can hope to see finally deviations in precision measurements occur typically for two reasons.

If the new physics occurs in loop diagrams
their effect is usually suppressed by a loop factor $\alpha \slash {4 \pi}$
and very high precision is required to see it.
If the new physics is already
on the Born level but at very high masses
the effects are suppressed by propagator factor
$\frac{s}{s-m^2_{N_P}-\imath m_{N_P} \Gamma}$
so that is important to work at the highest possible energies.

Both effects have already been used successfully in the past.

For example, ten years ago LEP could predict the mass of the top from its loop effects.
It was found at TEVATRON with exactly such mass.
In the same way it is hoped that in ten years from now a linear collider can obtain effects of new physics.

Linear lepton colliders will provide the opportunity to
investigate photon collisions at energies and luminosities close
to these in $e^{+} e^{-}$ collisions \cite{gg_proposal}.

The possibility to transform the future linear
$e^{+}e^{-}$-colliders into the $\gamma \gamma$ and $\gamma e$
-colliders with approximately the same energies and luminosities
was shown. The basic $e^+ e^-$-colliders can be transformed into
the $e \gamma$- or $\gamma \gamma$-colliders. The intense $\gamma$
-beams for photon colliders are suggested to be obtained by
Compton scattering of laser lights which is focused on electrons
beams of basic $e^+ e^-$-accelerators.

The electron and photon linear colliders of next generation will attack
unexplored higher energy region where new behaviour can turn up.
In this area the photon colliders have a number of advantages.

-- The first of the above advantages is connected with
the better signal/background ratio at both $e^+e^-$- and
$e\gamma\slash\gamma\gamma$-colliders in comparison with hadron ones.

-- The production cross sections at photon colliders are usually larger
than those at electron colliders.

-- The photon colliders permit to investigate both of the problems
of new physics and those ones of "classical" hadron physics and QCD.

So it is exclusively important task
to use possibilities of $\gamma\gamma$-colliders to realize the
experiments of the next generation.

If a light Higgs exists one of the main tasks of a photon collider will be the measurement
of the partial width $\Gamma (H \to \gamma\gamma)$.
Not to be limited by the error from luminosity determination
the luminosity of the collider at the energy
of the Higgs mass has to be known with a precision of around $1 \%$.

To produce scalar Higgses the total angular momentum of the two photons
has to be $J\!=\!0$.
In this case the cross section
$\gamma\gamma \to l^{+} l^{-}$ is suppressed by factor $m^2_{l} \slash s$
and thus not usable for luminosity determination.

In the SM the couplings of the
gauge bosons and fermions are constrained
by the requirements of gauge symmetry.
In the electroweak sector this
leads to trilinear and quartic
interactions between the gauge bosons
with completely specified couplings.

The trilinear and quartic
gauge boson couplings probe different
aspects of the weak interactions.
The trilinear couplings directly test the
non-Abelian gauge structure, and possible
deviations from the SM forms have been
extensively studied. In contrast,
the quartic couplings can be
regarded as a more direct way
of consideration of electroweak symmetry
breaking or, more generally, on
new physics which couples to electroweak
bosons.

In this respect it is quite possible
that the quartic couplings deviate
from their SM values while the
triple gauge vertices do not.
For example, if the mechanism
for electroweak symmetry breaking
doesn't reveal itself through the discovery
of new particles such as the
Higgs boson, supersymmetric particles
or technipions it is possible
that anomalous quartic couplings
could provide the first evidence
of new physics in this sector
of electroweak theory.

The production of several vector bosons
is the best place to search directly
for any anomalous behaviour of
triple and quartic couplings.

We wait for colliders
with higher center of mass energy in
order to produce a final state with
three or more gauge bosons and to test
the quartic gauge-bosons couplings. The
future project TESLA \cite{tdr} is the real
candidate for
investigation of process three boson
production.

Previously the cross sections for triple gauge bosons production
in framework of the SM presented for $e^{+}e^{-}$-colliders and
hadronic colliders.

By using of transforming a linear $e^{+}e^{-}$ collider in a
$\gamma\gamma$ collider, one can obtain very energetic photons
from an electron or positron beams. Such machines as TESLA which
will reach a center of mass energy $\sim 1000 GeV$ with high
luminosity ($\sim 10^{33} cm^{-2} s^{-1}$) will be able to study
multiple vector boson production with high statistics.

For obvious kinematic reasons,
processes where at least one
of the gauge bosons is a photon
have the largest cross sections.

We examine the production of
three vector boson in $\gamma\gamma$ collisions
through the reaction
\begin{eqnarray}
\gamma+\gamma \to W^{+} + W^{-} + Z^{0},
\end{eqnarray}
using beams of polarized photons.

1. This process involve only interactions between the gauge bosons
making more evident any deviations from predictions of the SM
gauge structure.

2.
There is no tree-level contribution involving
the Higgs boson which excludes all the
uncertainties coming from the scalar
sector.

We analyze the total cross section of the process of interaction
of the two $\gamma$  with fixed polarization, as well as the
dynamical distributions of the final state vector bosons.

The measurement of cross section
and asymmetries of this process is
complimentary to the analysis of the production of vector boson
pairs.

So
the photon linear colliders have the great physical potential \cite{tdr, gg_other}
(Higgs and SUSY particles searching,
study of anomalous gauge boson couplings
and hadronic structure of photons etc.).
Performing of this set of investigations
requires a fine measurement of the luminosity of photon beams.
For this purpose some of the well-known
and precisely calculated reactions
(see, for example, $\gamma\gamma\to 2 f, 4 f$, \cite{gg_llg_excl, gg_ll_corr, gg_2f_denner, gg_4l_moretti, gg_4l_kapusta, gg_4l_minsk}) are traditionally used.

It was shown that
it is convenient to use the events of $\gamma\gamma\to l^+l^-$ process
for measuring the luminosity of the $J\!=\!2$-beams
($J$ is the total angular momentum of initial photon couple).
Here $l$ is the unpolarized light lepton ($e$ or $\mu$).
It is the dominating QED process on $J\!=\!2$ beams
and its events are easily detected.

The difficulties appear in the calibration of
photon beams of similar helicity
(the total helicity of $\gamma \gamma$-system $J\!=\!0$)
since the small magnitude of cross sections
of the most QED processes.
For example, the leading term of cross section of $\gamma \gamma \to l \bar{l}$
scattering on $J\!=\!0$-beams is of order $\alpha \slash \pi$ ($\approx 0.002$).

We have found that the exclusive reaction $\gamma\gamma\to l^+l^-\gamma$
provides the unique opportunity to measure the luminosity of $J\!=\!0$ beams
on a linear photon collider.

One of the main purposes of the linear photon collider
is the $s$-channel of the Higgs boson production
at energies about $\sqrt{s}=120 GeV$ \cite{tdr}.
That is the reason of using this value of c.m.s. energy in our analysis.

\section{ Two lepton production with photon in $\gamma\gamma$-colli\-sions }

The two various helicity
configuration of the $\gamma \gamma$-system leads to the different spectra
of final particles
and requires the two mechanisms of beam calibration.
We have analyzed \cite{gg_llg_excl} the behaviour of the
$\gamma\gamma\to l^+l^-\gamma$ reaction
on beams with various helicities
as a function of the parameters of detectors.
We have performed the detail comparison of cross section on $\gamma^+\gamma^+$-
($J=0$) and $\gamma^+\gamma^-$-beams  ($J=2$).
Since experimental beams are partially polarized
the ratio of cross sections of $\gamma \gamma \to l^+ l^- \gamma$ scattering on $J\!=\!0$ to $J\!=\!2$-beams
should be high
for the effective luminosity measurement.
We have outlined the conditions that greatly restrict the
observation of the process on "$J=2$" beams, remaining the "$J=0$" cross
section almost unchanged.

Finally we estimate the precision of luminosity measurement.

Consider the process
\begin{eqnarray}
\gamma(p_1, \lambda_1)+\gamma(p_2, \lambda_2) \to f(p_1{}', e_1{}') + \bar{f}(p_2{}', e_2{}') + \gamma(p_3, \lambda_3),
\end{eqnarray}
where $\lambda_{i}$ and $e_{i}{}'$ are the photon and the fermion helicities.

We denote the c.m.s. energy squared by $s = {\left(p_1+p_2\right)}^2 = 2 \spr{p_1}{p_2},$
the final-state photon energy by $w$.
For the differential cross section the normalized
final-state photon energy (c.m.s. is used) $x=w\slash\sqrt{s}$ is introduced.
The differential cross section ${d\sigma}\slash{d x}$
appears to be the energy spectrum of
final-state photons.

The matrix elements are obtained using two methods: the massless
helicity amplitudes \cite{ha} for the fast estimations and the exact
covariant analysis \cite{belarus} including finite fermion mass. Since
final-state polarizations will not be measured we have summarized
over all final particles helicities. The integration over the
phase space of final particles is performed numerically using the
Monte-Carlo method \cite{mc}.

The calculations have been performed for various experimental restrictions on
the parameters of final particles.
Events are not detected if energies and angles are below the corresponding threshold values.
The considering restrictions on the phase-space of final particles (the cuts) are denotes as follows:

$\bullet$ Minimum final-state photon energy: $\omega_{cut}$,

$\bullet$ Minimum fermions energy: $E_{f,cut}$,

$\bullet$ Minimum angle between any final and any initial particles (polar angle cut): $\Theta_{cut}$,

$\bullet$ Minimum angle between any pair of final particles: $\varphi_{cut}$.

\begin{figure}[h!]
\leavevmode
\begin{minipage}[b]{.5\linewidth}
\centering
\includegraphics[width=\linewidth, height=3.9in, angle=0]{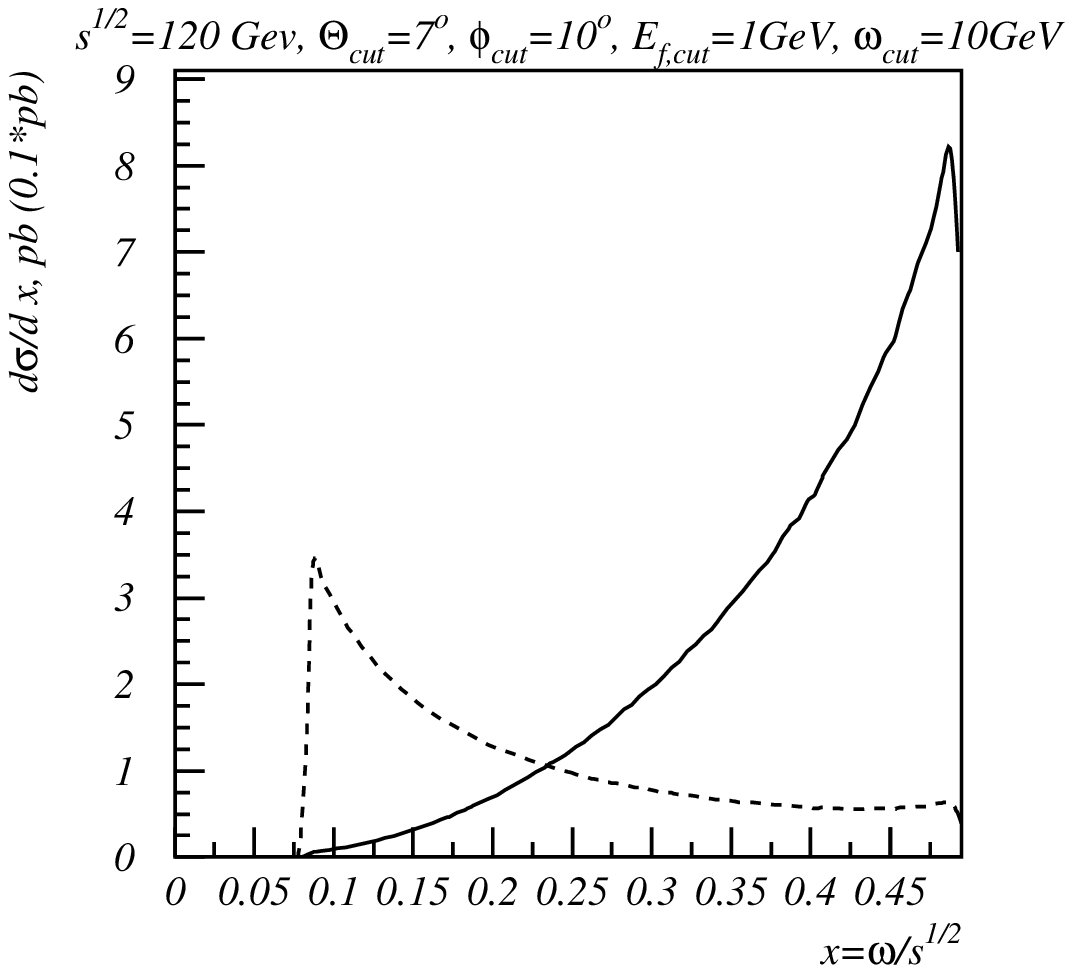}
\end{minipage}\hfill
\begin{minipage}[b]{.5\linewidth}
\centering
\includegraphics[width=\linewidth, height=3.9in, angle=0]{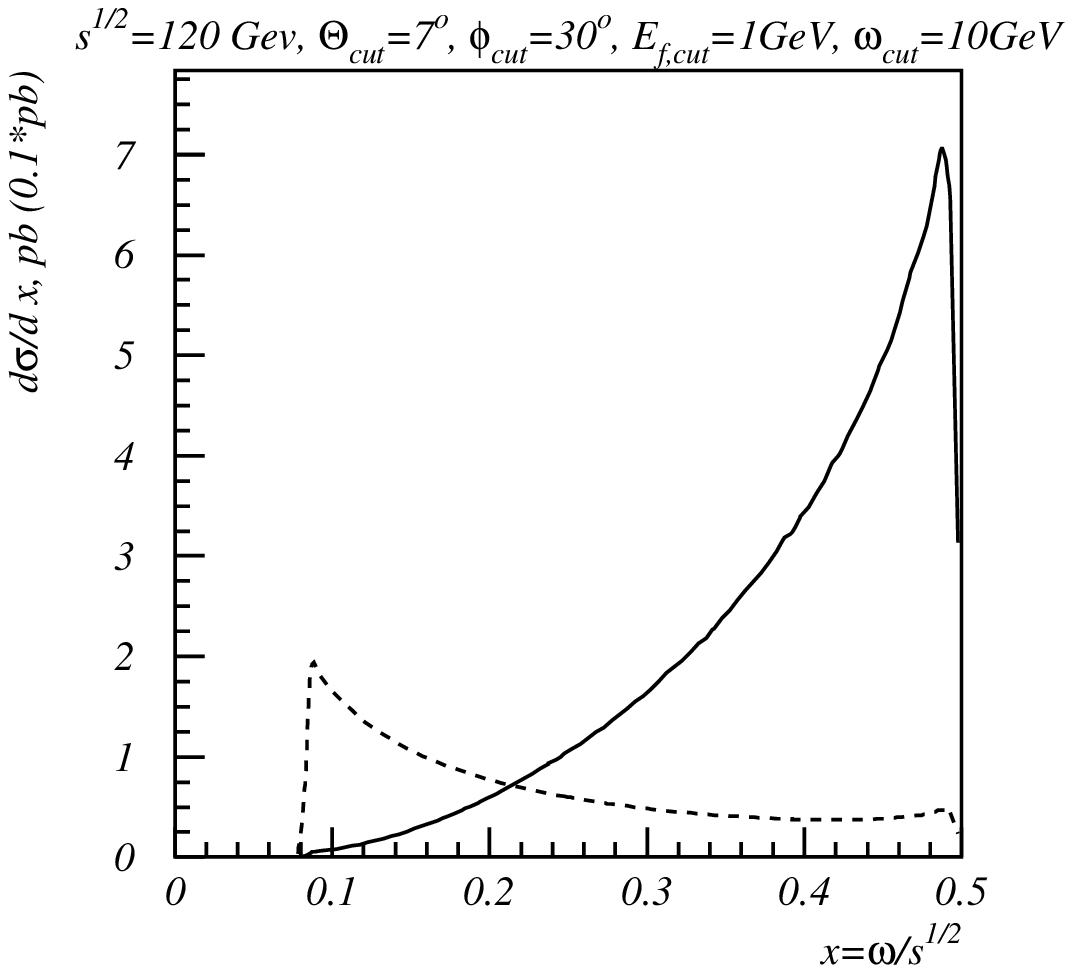}
\end{minipage}
\vspace{-20pt}
\caption{
Final-state photon energy spectrum for $J\!=\!0$ (solid) and $(J\!=\!2)*0.1$ (dotted)
at $\sqrt{s}=120GeV$ and cuts:
$\Theta_{min} \!=\! 7^o$, $\varphi_{min} \!=\! 10^o$ (left) and $\varphi_{min} \!=\! 30^o$ (right),
$E_{f, min} \!=\! 1 GeV$, $\omega_{min} \!=\! 10 GeV$.
}\label{diffs_llg}
\end{figure}

Consider the energy spectrum of final photons.
In fig. \ref{diffs_llg} the spectra for $J\!=\!0$ and $(J\!=\!2)$ are presented
(the $(J\!=\!2)$-cross section is scaled on factor $0.1$ for the convenience).
The differential cross section
${d \sigma}\slash{d x}$ on $J=2$ beams
decreases while one on $J=0$ beams raises with increasing of the
final-state photon energy.
This leads to the conclusion that if one increases the threshold
on $w$,
the process on $J\!=\!2$ beams will be greatly restricted,
but the rate of $\!J=0\!$ events remains almost unchanged.

\begin{figure}[h!]
\leavevmode
\begin{minipage}[b]{.5\linewidth}
\centering
\includegraphics[width=\linewidth, height=4.0in, angle=0]{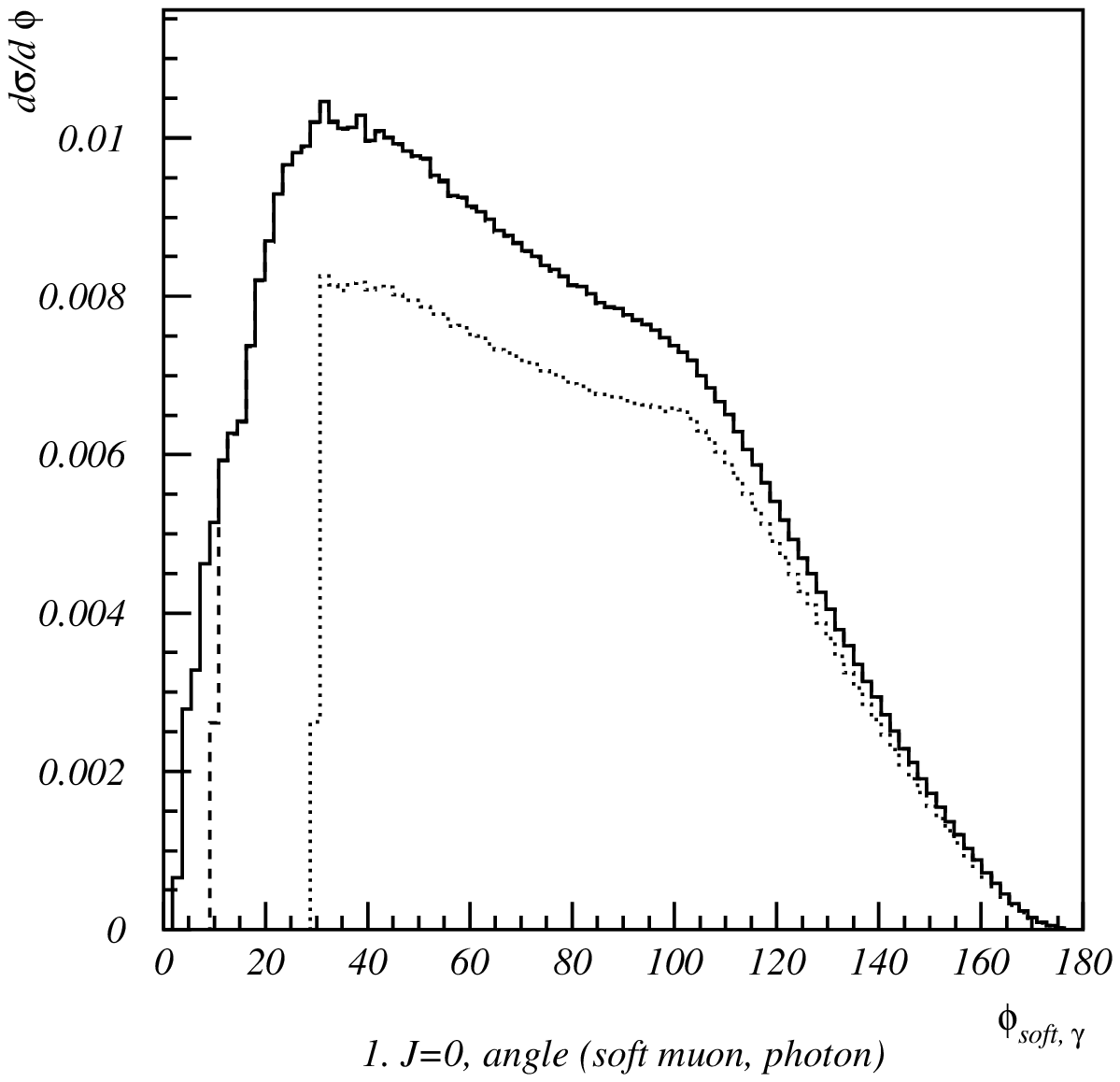}
\end{minipage}\hfill
\begin{minipage}[b]{.5\linewidth}
\centering
\includegraphics[width=\linewidth, height=4.0in, angle=0]{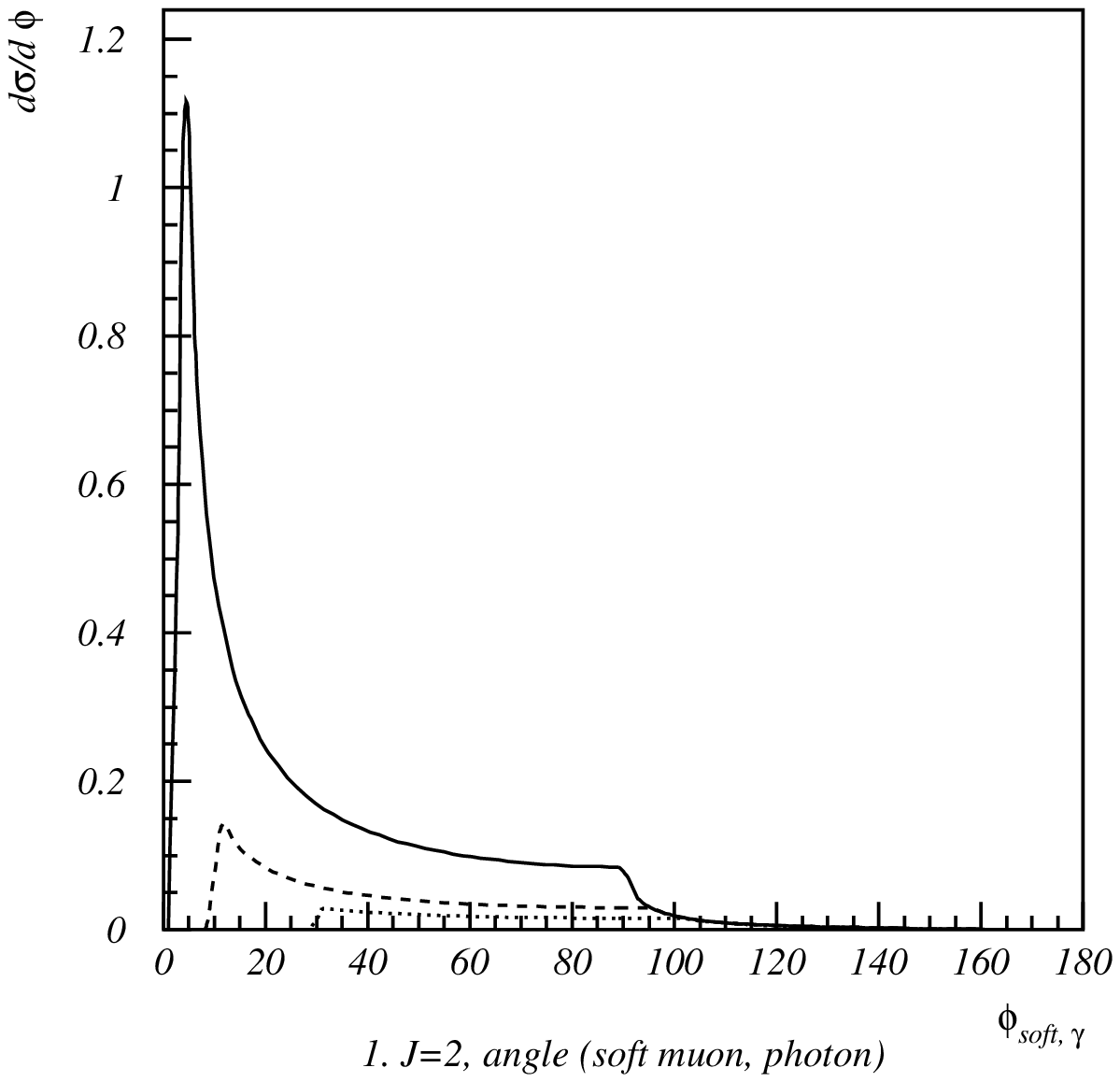}
\end{minipage}
\caption{
Angular plots (angle between fermion with lowest energy and final photon)
at $\sqrt{s}=120GeV$
at various sets of cuts:
$\bullet$ $\Theta_{min} = 7^o$, $\varphi_{min} = 3^o$, $E_{f, min}=1 GeV$, $\omega_{min} = 1 GeV$ (solid line);
$\bullet$ $\Theta_{min} = 7^o$, $\varphi_{min} = 10^o$, $E_{f, min}=1 GeV$, $\omega_{min} = 10 GeV$ (dashed line);
$\bullet$ $\Theta_{min} = 7^o$, $\varphi_{min} = 30^o$, $E_{f, min}=5 GeV$, $\omega_{min} = 20 GeV$ (dotted line).
}\label{ang_llg}
\end{figure}

\begin{figure}[h!]
\centering
\leavevmode
\begin{minipage}[b]{.33\linewidth}
\centering
\includegraphics[width=\linewidth, height=3.2in, angle=0]{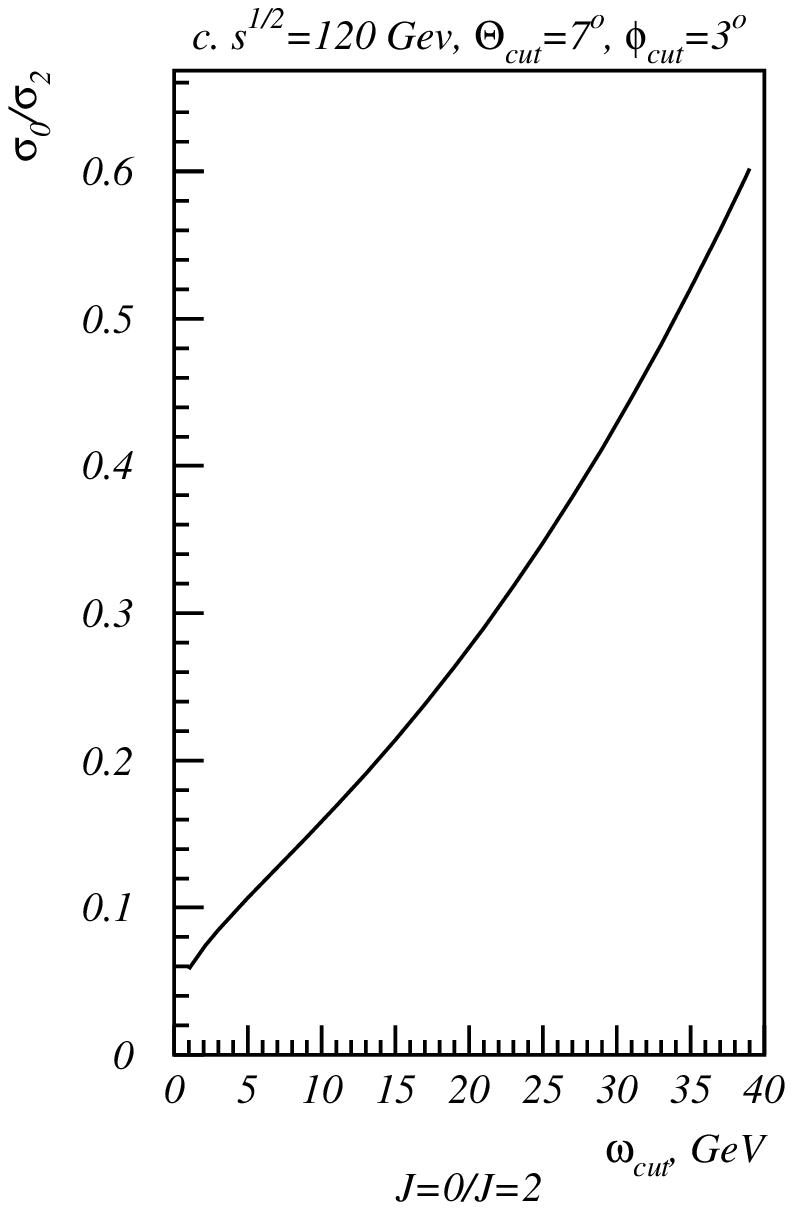}
\end{minipage}\hfill
\begin{minipage}[b]{.33\linewidth}
\centering
\includegraphics[width=\linewidth, height=3.2in, angle=0]{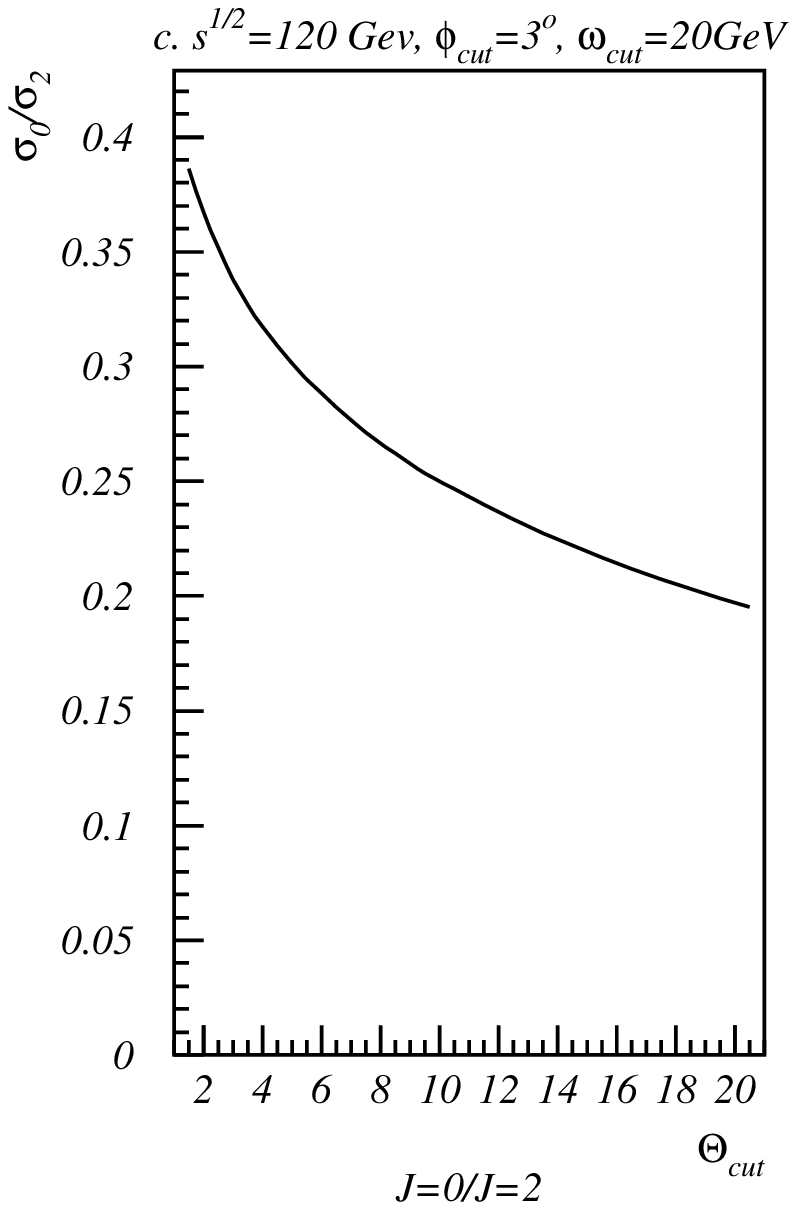}
\end{minipage}\hfill
\begin{minipage}[b]{.33\linewidth}
\centering
\includegraphics[width=\linewidth, height=3.2in, angle=0]{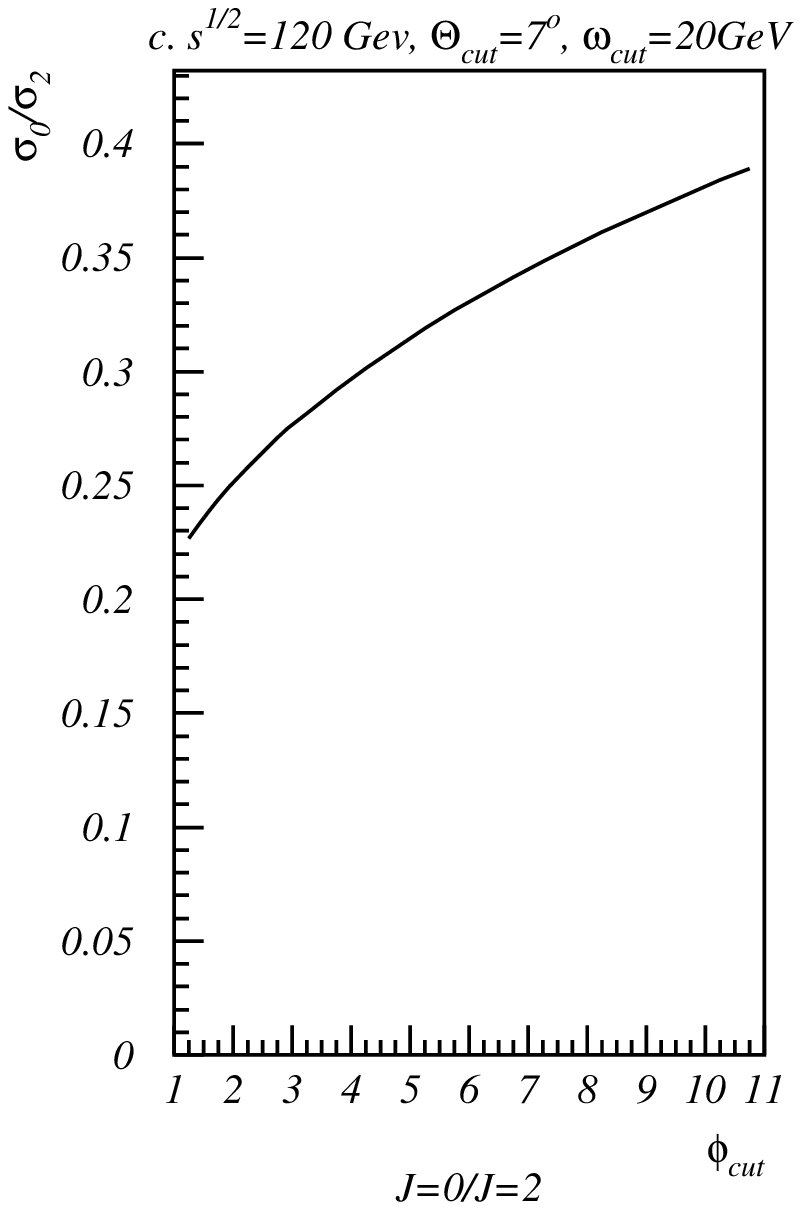}
\end{minipage}
\caption{
Ratios of total cross sections on $J\!=\!0$ and $J\!=\!2$ beams
depending on different cuts.
}\label{ratios_llg}
\end{figure}

Next we analyze the total cross section dependence on the angular cuts
$\Theta_{min}$ and $\varphi_{min}$ (see figs. \ref{ang_llg}, \ref{ratios_llg}).
In $\gamma^+\gamma^+$- experiments ($J\!=\!0$) the most of
final fermions are radiated closely to the axis of initial photon
(polar axis) and can't be detected.
The $\gamma^+\gamma^-$- experiments
($J\!=\!2$) have the large fraction of particles emitted at large
angles. But they are likely emitted as the collinear
fermion-photon pairs, that also can't be separated in the
detector.
The angular spectra of the final particles is represented in fig. \ref{ang_llg}.
It shows that the rise of threshold on angle between final particles (up to $20^{o}$)
restricts the $J\!=\!2$-cross section,
while the $J\!=\!0$ reaction is almost unaffected.

It is clear that the ratio of these cross sections can be increased
by reducing the polar angle cut $\Theta_{min}$
and by raising the collinear angle cut $\varphi_{min}$.

In fig. \ref{ratios_llg} we present the
ratios of total cross sections  $\sigma_{J=0} \slash \sigma_{J=2}$
depending on various of experimental cuts.
Using also the graphs for spectra of final particles one can achieve
the ratio $\sigma_{J=0} \slash \sigma_{J=2}$ up to $1$
without sufficient decreasing of $\sigma_{J=2}$.

We have discovered
the ratio of events on $J\!=\!0$ and $J\!=\!2$ beams
strongly depends on the experimental cuts.
We obtained the region (the configuration of cuts)
where the processes on the both $J\!=\!0$ and $J\!=\!2$ beams
have the cross sections close by each other.
That is the region of small polar angle cut,
high collinear angle cut
as well as high minimal energy of final-state photons.
At these parameters the total cross sections
of $ \gamma\gamma\to f \bar{f}\gamma$ in
experiments using $\gamma^+\gamma^+$- and $\gamma^+\gamma^-$- beams
appear to be the same order of magnitude.

\begin{figure}[h!]
\leavevmode
\begin{minipage}[h]{.5\linewidth}
\centering
\includegraphics[width=\linewidth, height=3.5in, angle=0]{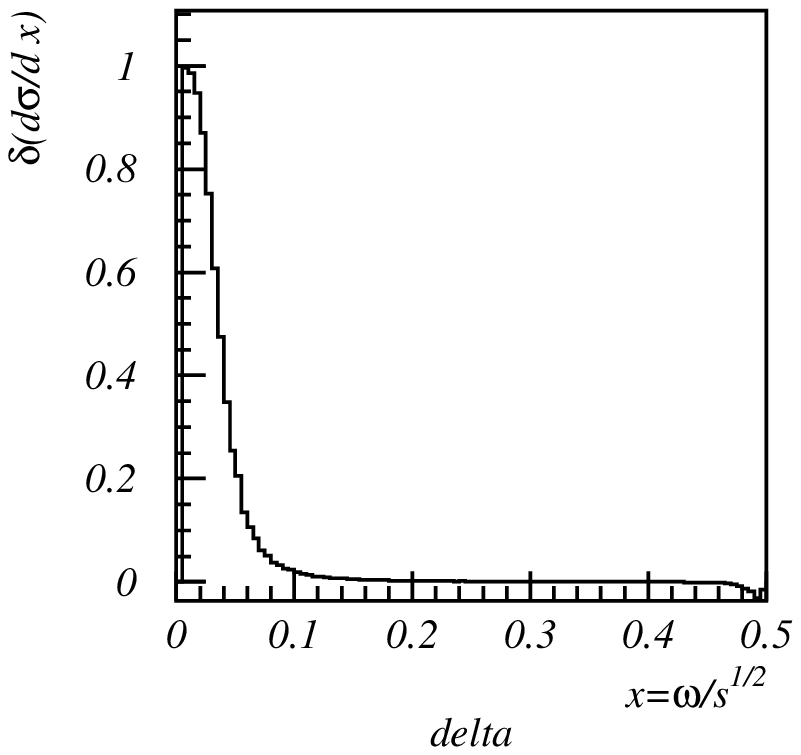}
\end{minipage}\hfill
\begin{minipage}[h]{.5\linewidth}
\centering
\includegraphics[width=\linewidth, height=3.5in, angle=0]{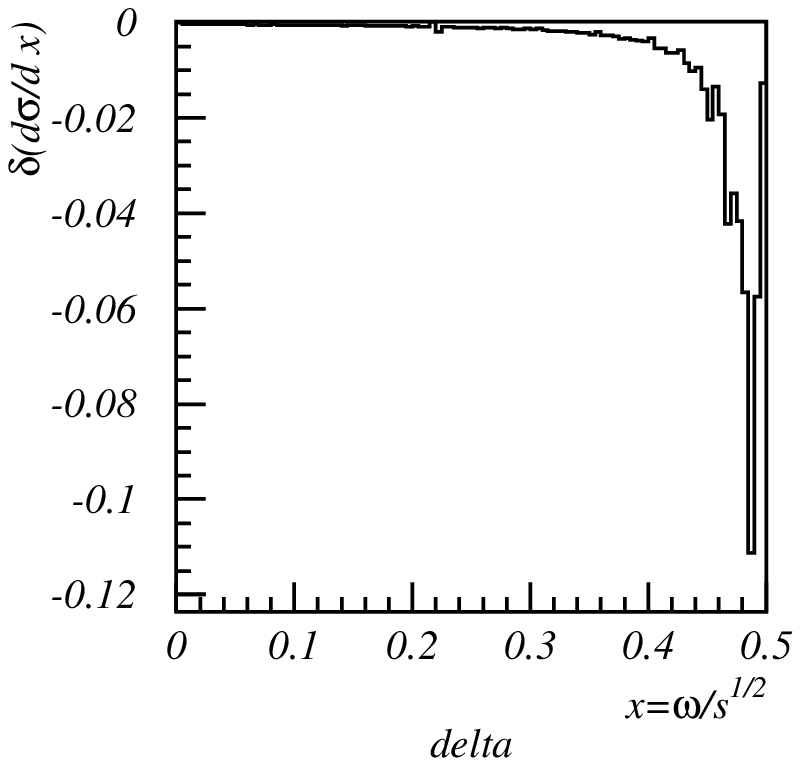}
\end{minipage}
\caption{
The relative mass contribution to
energy spectra of final photon
for $J\!=\!0$ (left) and $J\!=\!2$ (right) beams
($w_{cut}\!=\!1GeV$, $\E_{cut}\!=\!1GeV$, $\Theta_{cut}\!=\!7^o$, $\varphi_{cut}\!=\!3^o$).
}\label{mass_llg}
\end{figure}

The mass contribution is small
in the great part of phase space of final particles.
The most significant contribution is for the $J\!=\!0$ energy spectra
(see fig. \ref{mass_llg}).
The high value of the contribution corresponds
to regions where the differential cross section is minimal.
The mass contribution to the total cross section is below the $1\%$ level
at any realistic set of cuts.
It means that the helicity amplitudes is a good approach
for study the $\gamma\gamma\to l^+l^- \gamma$ process.

\subsection{Luminosity measurement of $J\!=\!0$ beams.}

Finally, we analyze the precision of luminosity measurement \cite{gg_llg_excl} that
can be achieved using the reaction $\gamma\gamma\to f \bar{f}
\gamma$.

The most interest are offered by the two kinds of measurement.
The first one is the measuring of beams luminosity with the wide energy spectrum.
The second one is the same measurement for the narrow band around the energy of Higgs boson production.

We use for consideration the following TESLA project parameters \cite{tdr}:

1. luminosity
\begin{eqnarray*}
{\cal L} (\sqrt{s'} > 0.8 \sqrt{s'_{\rm max}}) & = & 5.3 \cdot 10^{33} {cm}^{-2}s^{-1},\\
{\cal L} (m_H \pm 1 GeV) & = & 3.8 \cdot 10^{32} {cm}^{-2}s^{-1};
\end{eqnarray*}

2. polarization ${\cal P} \approx 90\%$.

Our calculations allow to
choose the set of cuts with the high $J\!=\!0$ cross section
and high ratio $\sigma_{J\!=\!0}\slash \sigma_{J\!=\!2}$:
$\omega_{cut} \!=\! 20 GeV$,
$E_{f,cut} \!=\! 5 GeV$,
$\Theta_{cut} \!=\! 6^\circ$,
$\varphi_{cut} \!=\! 30^\circ$.
For these cuts the total cross sections have the following values:
\begin{eqnarray*}
\sigma (J=0) & = & 0.82 pb,\\
\sigma (J=2) & = & 1.89 pb.
\end{eqnarray*}

So for the precision of luminosity measurement
in a 2 years run ($2 \cdot 10^7 {\rm s}$) one can obtain:
\begin{eqnarray*}
\frac{\Delta {\cal L}}{{\cal L}}
\left(\sqrt{s'} > 0.8 \sqrt{s'_{\rm max}}\right)
& = & 0.35 \%, \\
\frac{\Delta {\cal L}}{{\cal L}}
\left(m_H \pm 1GeV\right)
& = & 1.3 \%.
\end{eqnarray*}

\section{ Lepton-antilepton production in high energy polarized photons interaction }

The luminosity measurement at $J\!=\!2$ beams will be performed
using the reaction $\gamma\gamma\to l^+l^-$. It has the great
cross section that provides the number of events enough for the
$0.1\%$ precision of luminosity determination.

The main task is to calculate the cross section with maximal
precision. For realization of this purpose we have calculated the
complete one-loop QED radiative corrections to cross section of
$\gamma\gamma\to l^+l^-$ process including the hard photon
bremsstrahlung.

The major feature of $\gamma\gamma\to f \bar{f}$ process is the small value of
cross section if the total angular momentum of $\gamma\gamma-$beams equals zero.

We analyze both the angular spectra and the invariant
distributions of final particles. The angular spectrum of final
leptons is calculated in form $d \sigma \slash d
\cos{\Theta(p_{l},p_{\gamma})}$. It is more convenient to use
Lorentz-invariant results for the experimental reasons. Therefore
we analyze the process $\gamma\gamma\to f\bar{f}[+\gamma]$
including $O(\alpha)$-corrections using the method of covariant calculations \cite{belarus}. The
invariant differential cross section is calculated in the form $d
\sigma \slash d (p_{l}-p_{\gamma})^2$ and can be used in the
arbitrary experimental configuration.

The detailed analysis of the $\gamma\gamma\to l^+l^-$ process
can be found in the adjacent report
in these proceedings \cite{gg_ll_corr}.

For the measurement the luminosity of $J\!=\!2$ beams
one will use the events of $\gamma\gamma\to l^+l^-$ process.
The precision of measurement the luminosity of $J\!=\!2$ beams
that can be achieved using $\gamma\gamma\to l^+l^-$ process
can be calculated in the same way that one for $J\!=\!0$ beams.
We introduce the $\omega_{max}$ parameter for
the maximal energy of bremsstrahlung photon
that will still result
the detection of single exclusive $\gamma\gamma\to l^+l^-$ event.
For the supposed TESLA detector parameters
($\omega_{max} \!=\! 1 GeV$,
$E_{f,cut} \!=\! 1 GeV$,
$\Theta_{cut} \!=\! 7^\circ$) one can obtain:

\begin{eqnarray*}
\frac{\Delta {\cal L}}{{\cal L}}
\left(\sqrt{s'} > 0.8 \sqrt{s'_{\rm max}}\right)
& = & 0.04 \%, \\
\frac{\Delta {\cal L}}{{\cal L}}
\left(m_H \pm 1GeV\right)
& = & 0.1 \%.
\end{eqnarray*}

The achieved precision is sufficient for the huge variety of
experiments at the photon collider.


\section{ Production of four leptons in $\gamma\gamma$ electroweak interaction }

Since the high accuracy
and relatively clean environment are provided by a linear collider,
a precision calculation of background including $\gamma\gamma \rightarrow
(...) \rightarrow 4l$ process is necessary.

Total cross sections of such  interactions  have been already
calculated and analyzed in refs. \cite{gg_4l_old} about 30 years
ago  and were found to be large enough:

 $$\sigma = 6500nb\;\;\;\;(\gamma\gamma
\rightarrow 2e^-2e^+),$$ $$\sigma = 5.7nb\;\;\;\;(\gamma\gamma
\rightarrow e^+e^-\mu^+\mu^-),$$ $$\sigma =
0.16nb\;\;\;\;(\gamma\gamma \rightarrow 2\mu^-2\mu^+).$$

\noindent
 However these calculations have  used the low energy approximation and
 obtained results are not applicable for modern high energy investigations.

There was applied the algorithm  ALPHA \cite{gg_4l_moretti} for
automatic computations of scattering amplitude.
However modern
high energy experiments require the cross section calculation and analysis at fixed
polarization states of initial and final particles that ALPHA
method couldn't provide.

Our purpose was investigation of process $\gamma\gamma\to 4l$
using parameters of linear colliders of new generation such as
TESLA \cite{tdr}.

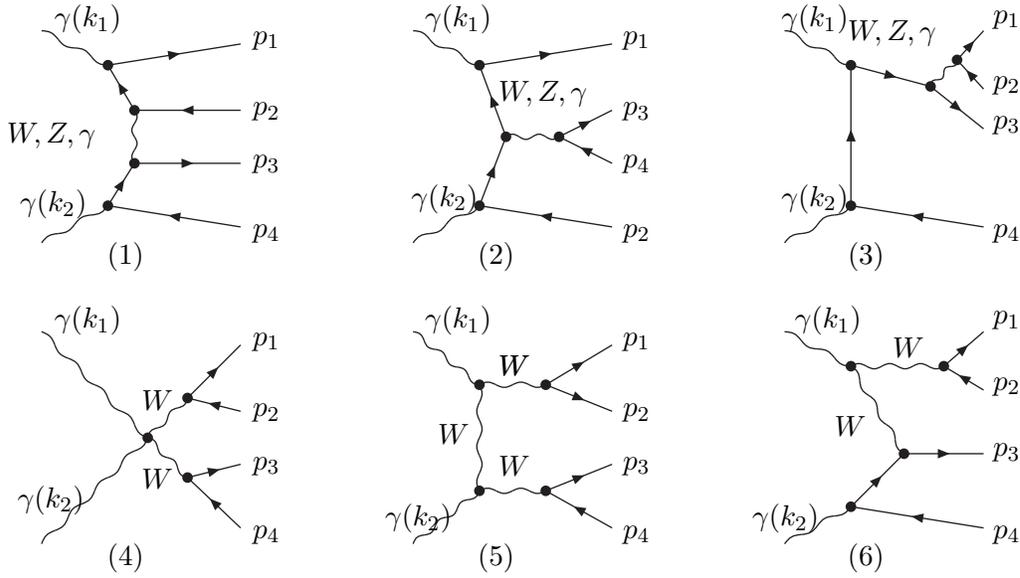
\begin{figure}[h!]
\begin{center}
\begin{picture}(400,100)
\SetOffset(0, -100)

\Photon(10,180)(35,167){1}{2} \Photon(10,100)(35,114){1}{2}
\ArrowLine(35,167)(85,175) \ArrowLine(85,106)(35,114)
\ArrowLine(45,150)(35,167) \ArrowLine(35,114)(45,130)
\Photon(45,150)(45,130){1}{2} \ArrowLine(85,150)(45,150)
\ArrowLine(45,130)(85,130) \Text(15,185)[l]{\small $\gamma(k_1)$}
\Text(2,115)[l]{\small $\gamma(k_2)$}\Text(90,177)[l]{\small
$p_1$}\Text(90,150)[l]{\small $p_2$} \Text(90,130)[l]{\small
$p_3$}\Text(90,104)[l]{\small $p_4$} \Text(35,95)[l]{\small
$(1)$}\Text(-3,141)[l]{\small
$W,Z,\gamma$}\Vertex(35,167){2}\Vertex(35,114){2}\Vertex(45,150){2}
\Vertex(45,130){2}

\Photon(150,180)(175,167){1}{2} \Photon(150,100)(175,114){1}{2}
\ArrowLine(175,167)(225,175) \ArrowLine(225,106)(175,114)
\ArrowLine(185,140)(175,167) \ArrowLine(175,114)(185,140)
\Photon(185,140)(205,140){1}{2} \ArrowLine(225,130)(205,140)
\ArrowLine(205,140)(225,150) \Text(155,185)[l]{\small
$\gamma(k_1)$} \Text(150,117)[l]{\small
$\gamma(k_2)$}\Text(230,177)[l]{\small
$p_1$}\Text(230,150)[l]{\small $p_3$} \Text(230,130)[l]{\small
$p_4$}\Text(230,104)[l]{\small $p_2$} \Text(175,95)[l]{\small
$(2)$}\Text(183,157)[l]{\small
$W,Z,\gamma$}\Vertex(175,167){2}\Vertex(175,114){2}\Vertex(205,140){2}
\Vertex(185,140){2}

\Photon(290,180)(315,167){1}{2} \Photon(290,100)(315,114){1}{2}
\ArrowLine(315,167)(345,159) \ArrowLine(365,106)(315,114)
\ArrowLine(315,114)(315,167) \ArrowLine(345,159)(365,143)
\Photon(345,159)(355,169){1}{2} \ArrowLine(365,158)(355,169)
\ArrowLine(355,169)(365,180) \Text(290,185)[l]{\small
$\gamma(k_1)$} \Text(290,117)[l]{\small
$\gamma(k_2)$}\Text(370,185)[l]{\small
$p_1$}\Text(370,160)[l]{\small $p_2$} \Text(370,145)[l]{\small
$p_3$}\Text(370,104)[l]{\small $p_4$} \Text(315,95)[l]{\small
$(3)$}\Text(315,180)[l]{\small
$W,Z,\gamma$}\Vertex(315,167){2}\Vertex(315,114){2}\Vertex(345,159){2}
\Vertex(355,169){2}

\end{picture}
\end{center}

\begin{center}
\begin{picture}(400,100)
\SetOffset(0, -100)

\Photon(10,180)(50,140){1}{4} \Photon(10,100)(50,140){1}{4}
\ArrowLine(65,155)(85,175) \ArrowLine(85,106)(65,125)
\Photon(50,140)(65,155){1}{2}\Photon(50,140)(65,125){1}{2}
\ArrowLine(85,150)(65,155) \ArrowLine(65,125)(85,130)
\Text(15,185)[l]{\small $\gamma(k_1)$} \Text(1,117)[l]{\small
$\gamma(k_2)$}\Text(90,177)[l]{\small
$p_1$}\Text(90,150)[l]{\small $p_2$} \Text(90,130)[l]{\small
$p_3$}\Text(90,104)[l]{\small $p_4$} \Text(35,95)[l]{\small
$(4)$}\Text(48,155)[l]{\small $W$}\Text(48,125)[l]{\small
$W$}\Vertex(50,140){2}\Vertex(65,155){2}\Vertex(65,125){2}

\Photon(150,180)(175,160){1}{3} \Photon(150,100)(175,120){1}{3}
\ArrowLine(200,160)(225,175) \ArrowLine(225,106)(200,120)
\Photon(175,120)(175,160){1}{3}
\Photon(175,120)(200,120){1}{2}\Photon(175,160)(200,160){1}{2}
\ArrowLine(200,120)(225,130) \ArrowLine(200,160)(225,150)
\Text(155,185)[l]{\small $\gamma(k_1)$} \Text(140,110)[l]{\small
$\gamma(k_2)$}\Text(230,177)[l]{\small
$p_1$}\Text(230,150)[l]{\small $p_2$} \Text(230,130)[l]{\small
$p_3$}\Text(230,104)[l]{\small $p_4$} \Text(175,95)[l]{\small
$(5)$}\Text(183,168)[l]{\small $W$}\Text(183,130)[l]{\small
$W$}\Text(183,168)[l]{\small $W$} \Text(160,142)[l]{\small $W$}
\Vertex(175,160){2}\Vertex(175,120){2}\Vertex(200,120){2}
\Vertex(200,160){2}

\Photon(290,180)(315,167){1}{2} \Photon(290,100)(315,114){1}{2}
\Photon(315,167)(350,167){1}{3} \ArrowLine(365,106)(315,114)
\ArrowLine(315,114)(335,134) \ArrowLine(335,134)(365,134)
\Photon(315,167)(335,134){1}{3} \ArrowLine(365,158)(350,167)
\ArrowLine(350,167)(365,180) \Text(295,185)[l]{\small
$\gamma(k_1)$} \Text(279,110)[l]{\small
$\gamma(k_2)$}\Text(370,185)[l]{\small
$p_1$}\Text(370,160)[l]{\small $p_2$} \Text(370,135)[l]{\small
$p_3$}\Text(370,104)[l]{\small $p_4$} \Text(315,95)[l]{\small
$(6)$}\Text(332,175)[l]{\small $W$}\Text(310,145)[l]{\small
$W$}\Vertex(315,167){2}\Vertex(315,114){2}\Vertex(350,167){2}
\Vertex(335,134){2}

\end{picture}
\end{center}
\caption{The Feynman diagrams of four lepton production in $\gamma\gamma$ electroweak interaction.}
\label{wanja_fig1}
\end{figure}

There are six topographically different Feynman diagrams
of electroweak interactions for
describing of
this process (see fig. \ref{wanja_fig1}). Using
C-, P- and crossing symmetries one can built 40 different
diagrams.

The diagrams  containing charged current exchange are excluded
because only processes with four charged leptons in the final state
are considered.

The corresponding cross section has the
form:
\begin{eqnarray}\label{w2}
\begin{array}{c}
{\displaystyle
\sigma = \frac{1}{4(k_1k_2)}\int|M|^2d\Gamma.
}
\end{array}
\end{eqnarray}

Here $M=\sum_{i=1}^{3}{M_{i}}$, $M_{i}$ are the matrix elements of remaining diagrams (1)-(3) in fig. \ref{wanja_fig1}
and $ d\Gamma $ is the phase space element.

Squared matrix element is constructed using
the method of helicity amplitudes \cite{ha}
as well as the method of precision covariant calculation \cite{belarus}.
The first one
allows to calculate matrix element directly for each
definite polarization state of initial and final particles.
Amplitude
constructed by this method contains only invariants
without any bispinor, so many difficulties at squaring and
numerical integration are excluded.

 The explicit form of all
matrix elements constructed using helicity amplitude method one
can found in ref. \cite{gg_4l_minsk}.

The method of precision covariant calculations allows to
obtain matrix element without any approximations
and was used for
verifying results in each  step of our construction and
calculation.

 For the investigation of total and differential cross section the
method of Monte-Carlo integration was applied. If two or
more produced particles propagate into  very close direction, the
square of matrix element becomes very large (so-called collinear
peaks problem).
For the achievement of high precision the
Monte-Carlo generator was adopted.
Despite regular distributions of
kinematic variables (such distributions usually applied into
Monte-Carlo generators) we have used  irregular one, which is very
close to matrix element behavior. This proximity can be achieved
by choosing of several free parameters available in the
distribution function. So the adopted method of Monte-Carlo integration
leads to results with very  small numerical error, nearly
($0.1\% - 0.2\%$).
The accuracy of the method of helicity amplitude was investigated.
It was
arrived at a decision
that the results of two discussed methods
have good agreement.

\begin{figure}[h!]
\leavevmode
\begin{minipage}[b]{.5\linewidth}
\centering
\includegraphics[width=\linewidth, height=7.5cm, angle=0]{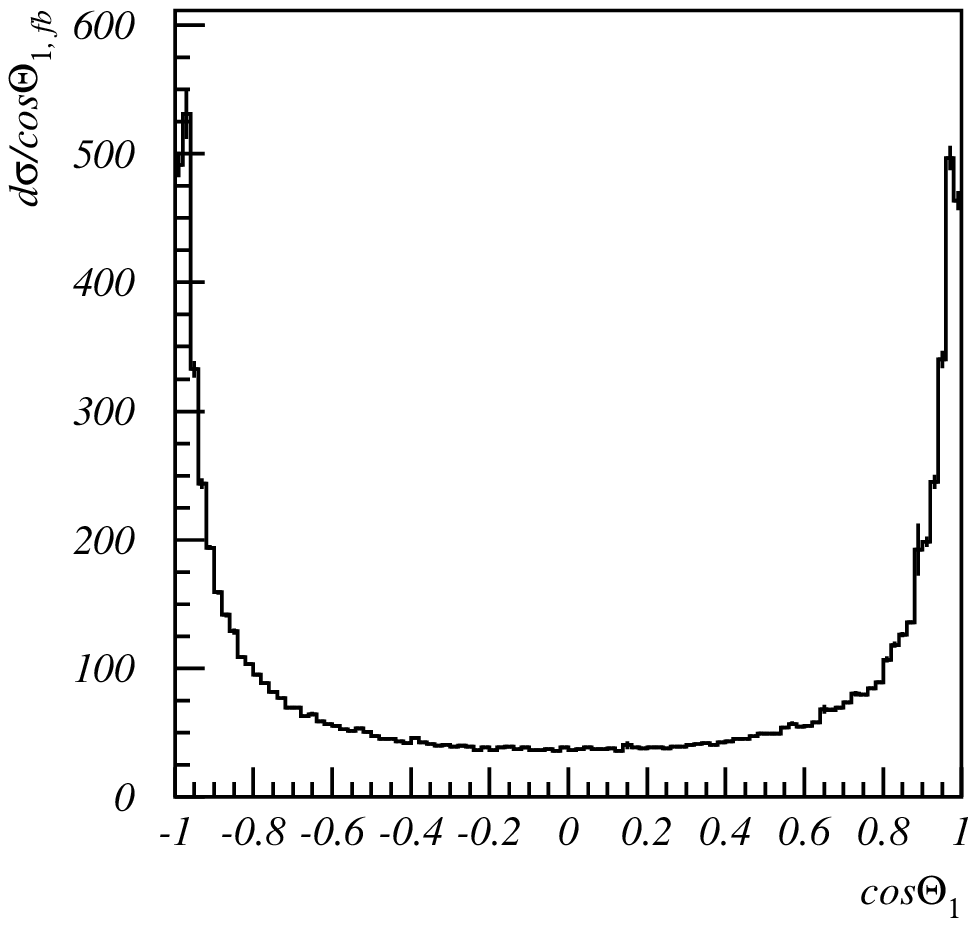}
\end{minipage}\hfill
\begin{minipage}[b]{.5\linewidth}
\centering
\includegraphics[width=\linewidth, height=7.5cm, angle=0]{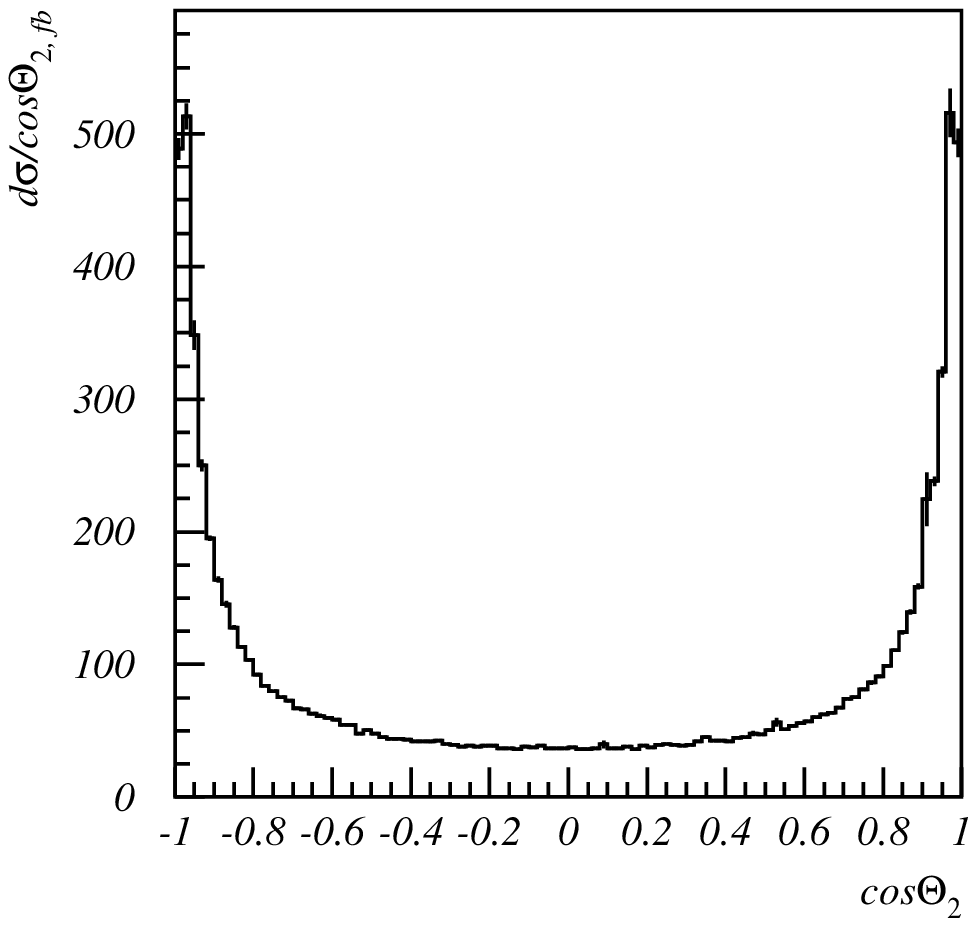}
\end{minipage}
\caption{
Spin averaged differential cross section of $\gamma\gamma \rightarrow 4l$
process at energy of interaction beams  $1\, TeV$ in c.m.s.
$\theta_1$ ($\theta_2$) is the angle between directions of the first (second) photon and the first lepton.
The magnitude of polar angle cut is equal to $7^o$.
}\label{wanja_fig2}
\end{figure}

\begin{figure}[h!]
\leavevmode
\begin{minipage}[b]{.5\linewidth}
\centering
\includegraphics[width=\linewidth, height=7.5cm, angle=0]{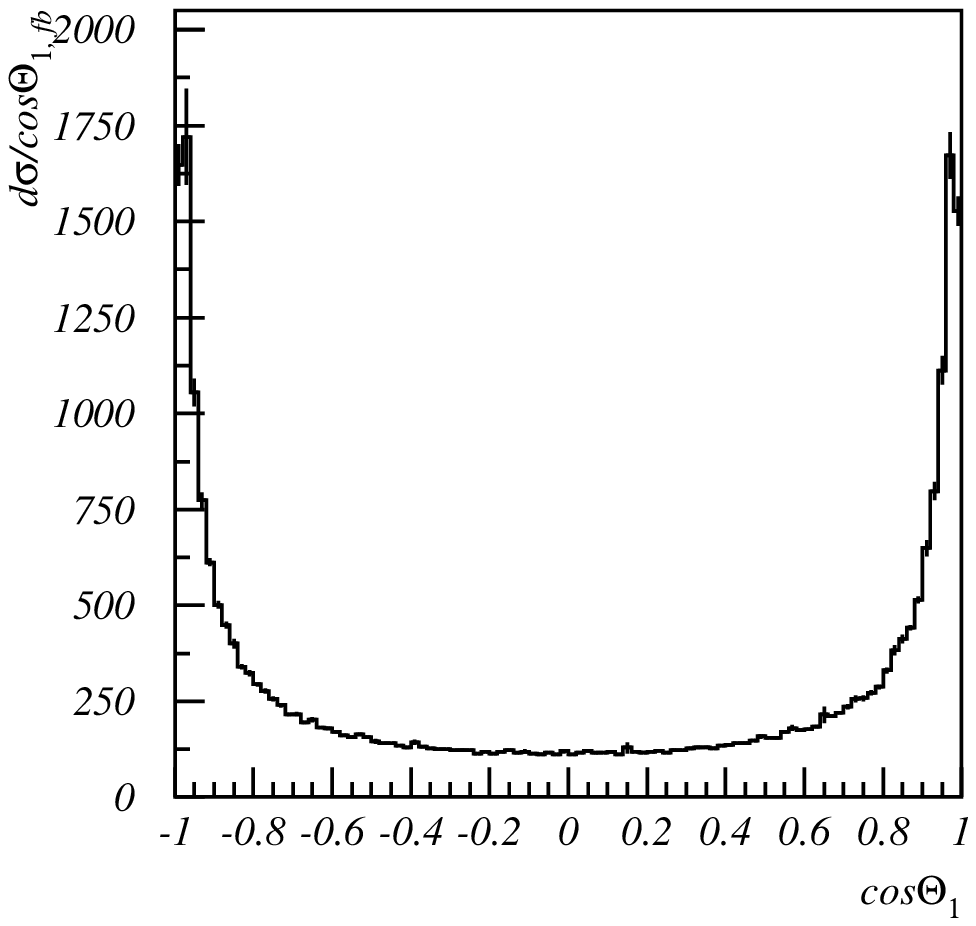}
\end{minipage}\hfill
\begin{minipage}[b]{.5\linewidth}
\centering
\includegraphics[width=\linewidth, height=7.5cm, angle=0]{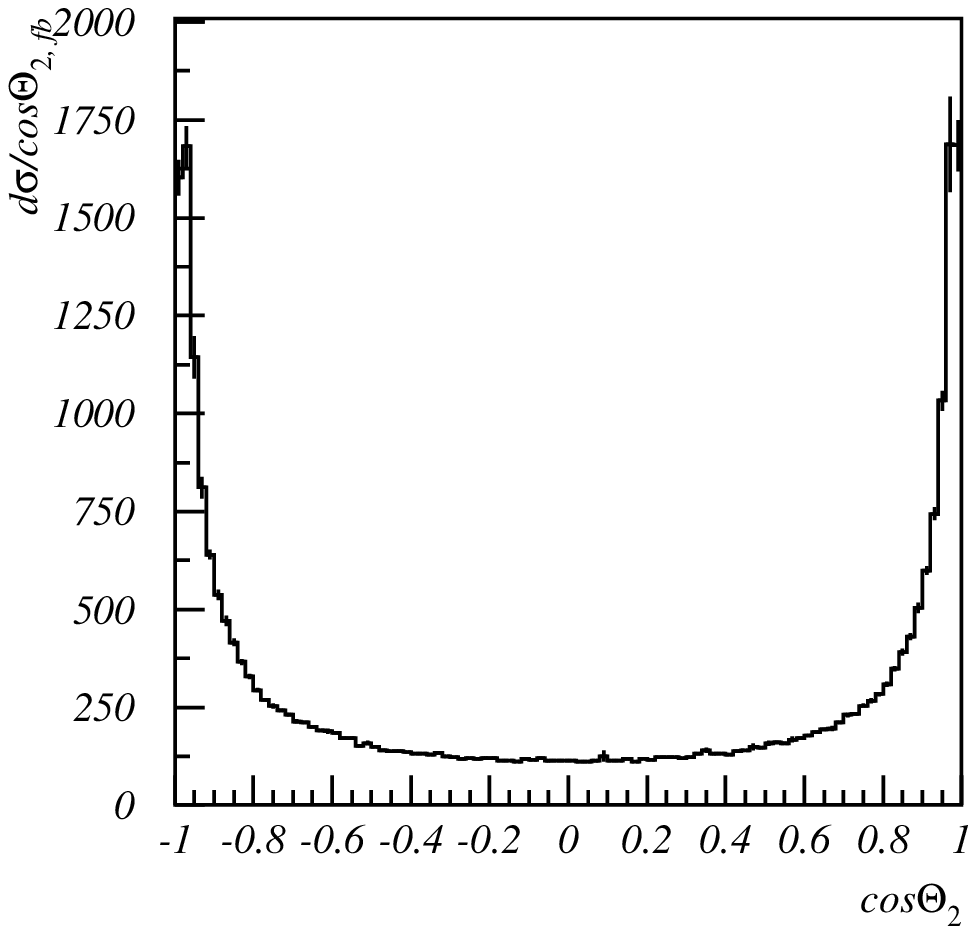}
\end{minipage}
\caption{
Spin averaged differential cross section of $\gamma\gamma \rightarrow 4l$ process
at energy of interaction beams  $500\, GeV$ in c.m.s.
$\theta_1$ ($\theta_2$) is the angle between directions of the first (second) photon and the first lepton.
The magnitude of polar angle cut is equal to $7^o$.
}\label{wanja_fig3}
\end{figure}

Some results, obtained for spin averaged differential cross section of electromagnetic $\gamma\gamma$- interaction,
can be seen in figs. \ref{wanja_fig2} and \ref{wanja_fig3}.
As it was expected, the total and differential cross sections have very strong dependence upon
kinematic cuts on polar angle (the angle between directions of initial and final particles).
We have found the magnitude of differential cross section increases significantly at polar angle tends to $0$ or $\pi$.
The cross section raises with decreasing of interaction particles energy.
It has symmetric (asymmetric) form in case of similar (different) spin configuration of scattering photons.
The cross section dependence on angle between initial photon momentum
and a set of final particles directions
has the same form due to electromagnetic interaction
results are only presented.



\section{The SM and anomalous amplitudes of $\gamma\gamma\rightarrow W^+W^-$}

Future high-energy linear $e^+e^-$-colliders in
$\gamma\gamma$-mode could be a very useful instrument to explore
the mechanism of symmetry breaking in electroweak interaction.
$\gamma\gamma$-colliders give the opportunity for mesuarement of a
light Higgs productions on resonance and $H\gamma\gamma$ coupling
because of the processes $\gamma\gamma\rightarrow H\rightarrow
ZZ,b\bar{b},WW$. In the case of heavy mass Higgs bosons (over
$1\,TeV$) a huge background from $\gamma\gamma\rightarrow
W^+_TW^-_T$ complicates a search of the Higgs signal.

Another way to probe the symmetry breaking is  examination of the
self couplings of the $W,Z$ bosons in non-minimal gauge models.
Since the cross section of $\gamma\gamma\rightarrow W^+W^-$ is
large the $WW$-production would be provided mainly by
$\gamma\gamma$ scattering \cite{igor_ref5}. So the Born cross section
$\sigma(\gamma\gamma\rightarrow WW)$ is about $110pb$ at $1\,TeV$
on unpolarized $\gamma$-beams at that all contributions of
different polarization sets of $W$-bosons are considered.
Corresponding cross section of $WW$-production in $e^+e^-$
collisions is an order of magnitude smaller and amount to $10pb$
at the same circumstances. In addition to the
$\gamma\gamma\rightarrow W^+W^-$ process one needs to consider a
reaction of $\gamma\gamma\rightarrow W^+W^-Z$ because of its cross
section becomes about  $5\%$ -- $10\%$ of the
$\sigma(\gamma\gamma\rightarrow WW)$ at energies $\sqrt{s}\geq
500\, GeV$. The anomalous tri-linear $\gamma WW$, $ZWW$ \cite{igor_ref4}
and quartic $\gamma\gamma WW$, $\gamma ZWW$, $ZZWW$ \cite{igor_ref6}
couplings induce deviations of Born cross sections
$\sigma(\gamma\gamma\rightarrow WW)$ and
$\sigma(\gamma\gamma\rightarrow WWZ)$ from Standard Model
values.

In order to evaluate contributions of anomalous couplings a cross section of  $\gamma\gamma\rightarrow W^+W^-$ must be
calculated with a high precision. Therefore one needs to calculate the main contribution of high order electroweak  effects:
one-loop correction, real photon and $Z$-boson emission.

The amplitude of $\gamma\gamma\rightarrow W^+W^-$ are defined as follows
\begin{align}\label{c1}
M = & G_v \epsilon_{\mu}(k_1)\epsilon_{\nu}(k_2)
\epsilon_{\alpha}(p_{+})\epsilon_{\beta}(p_{-})
M_T^{\mu\nu\alpha\beta},
\end{align}
where $k_1$, $k_2$, $p_{+}$, $p_{-}$ are $4$-momenta of the
$\gamma$, $\gamma$, $W^{+}$, $W^{-}$ respectively, and
$\epsilon_{\mu}(k_1),$ $\epsilon_{\nu}(k_2),$
$\epsilon_{\alpha}(p_{+}),$ $\epsilon_{\beta}(p_{-})$ -- 4-vectors
of polarizations accordingly. The constant $G_v$ is equal to $e^3
\cot\theta_W$. The total amplitude $M_T$ is defined by amplitudes
$M_i$ $(i=1,3)$ corresponding to the Feynman diagrams, presented in
fig. \ref{f1}:
\begin{align}\label{c2}
M_T^{\mu\nu\alpha\beta} = \sum_{i=1}^{3}, M_i^{\mu\nu\alpha\beta}.
\end{align}
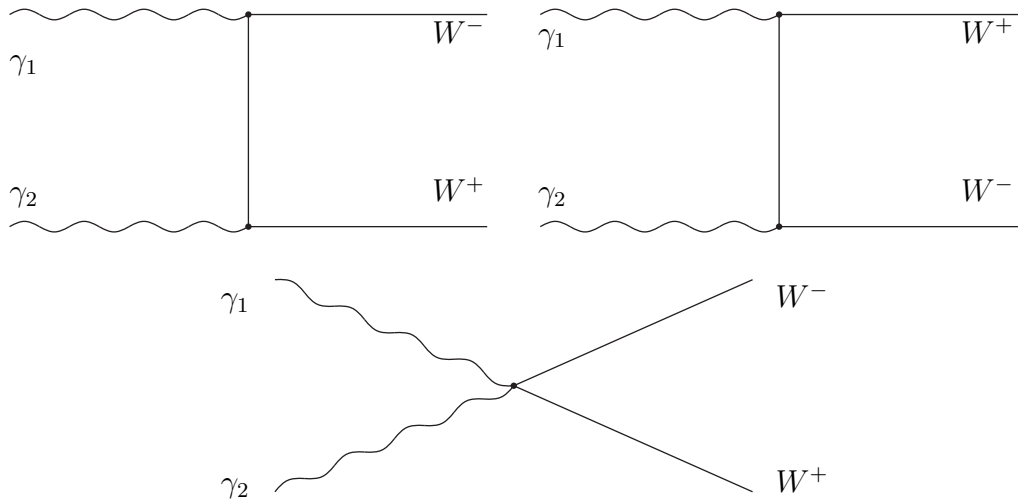
\begin{figure}[h]
\begin{center}
\begin{picture}(400,200)(0,0)
\Photon(10,110)(100,110){2}{4}\put(10,120){$\gamma_2$}
\Vertex(100,110){1.2}
\Line(100,110)(190,110)\put(170,120){$W^+$}
\Line(100,110)(100,190)
\Photon(10,190)(100,190){2}{4}\put(10,170){$\gamma_1$}
\Vertex(100,190){1.2}
\Line(100,190)(190,190)\put(170,180){$W^-$}
\Photon(210,110)(300,110){2}{4}\put(210,120){$\gamma_2$}
\Vertex(300,110){1.2}
\Line(300,110)(390,110)\put(370,120){$W^-$}
\Line(300,110)(300,190)
\Photon(210,190)(300,190){2}{4}\put(210,180){$\gamma_1$}
\Vertex(300,190){1.2}
\Line(300,190)(390,190)\put(370,180){$W^+$}
\Photon(110,10)(200,50){2}{4}\put(90,10){$\gamma_2$}
\Vertex(200,50){1.2}
\Photon(110,90)(200,50){2}{4}\put(90,80){$\gamma_1$}
\Line(200,50)(290,10)\put(300,10){$W^+$}
\Line(200,50)(290,90)\put(300,80){$W^-$}
\end{picture}
\end{center}
\caption{The Feynman diagrams for $W^+W^-$-production}\label{f1}
\end{figure}
These three amplitudes are constructed from a trilinear boson
${\Gamma}_{3}^{\mu\alpha\beta}$, a quartic boson
${\Gamma}_{4}^{\mu\nu\alpha\beta}$ vertices and a boson propagator
$D_{\alpha\beta}(p)$, where $p$ is 4-momentum of a virtual
$W$-boson.
\begin{align}\label{c3}
M_{1}^{\mu\nu\alpha\beta} = & {\Gamma}_{3}^{\mu\alpha\xi}
(-k_1,p_{+}) D_{\xi\lambda}(p_+ -
k_1){\Gamma}_{3}^{\nu\beta\lambda}(-k_2,p_-),
\end{align}
\begin{align}\label{c4}
M_{2}^{\mu\nu\alpha\beta} = & {\Gamma}_{3}^{\mu\beta\xi}
(-k_1,p_{-}) D_{\xi\lambda}(p_- -
k_1){\Gamma}_{3}^{\nu\alpha\lambda}(-k_2,p_+),
\end{align}
\begin{align}\label{c5}
M_{3}^{\mu\nu\alpha\beta} = & {\Gamma}_{4}^{\mu\nu\alpha\beta}.
\end{align}

Consider the anomalous quartic boson vertices only.
For this purpose the following $6$-dimensional $SU(2)_c$
Lagrangians \cite{igor_ref2, igor_ref1} have been chosen
\begin{eqnarray}\label{c9}
\begin{array}{c}
\displaystyle {\cal L}_0 = -\frac{e^2}{16\Lambda^2}a_0F^{\mu\nu}
F_{\mu\nu}\bar{W}^{\alpha}\bar{W}_{\alpha}, \\  \\
\displaystyle {\cal L}_c = -\frac{e^2}{16\Lambda^2}a_cF^{\mu\alpha}
F_{\mu\beta}\bar{W}^{\beta}\bar{W}^{\alpha}, \\  \\
\displaystyle \tilde{\cal L}_0 = -\frac{e^2}{16\Lambda^2}\tilde{a}_0
F^{\mu\alpha}\tilde{F}_{\mu\beta}\bar{W}^{\beta}\bar{W}^{\alpha}, \\
\end{array}
\end{eqnarray}
where the triplet of gauge bosons $\bar{W}_{\mu}$
\begin{eqnarray}\label{c10}
\begin{array}{c}
\displaystyle \bar{W}_{\mu} = \left(\frac{1}{\sqrt{2}}(W^+_{\mu}+
W^-_{\mu}),\frac{i}{\sqrt{2}}(W^+_{\mu}-W^-_{\mu}),\frac{1}
{\cos{\theta_W}}Z_{\mu}\right)
\end{array}
\end{eqnarray}
and the field-strength tensors
\begin{eqnarray}\label{c11}
\begin{array}{c}
\displaystyle F_{\mu\nu} = \partial_{\mu}A_{\nu}
-   \partial_{\nu}A_{\mu},
\quad 
\displaystyle W_{\mu\nu}^i = \partial_{\mu}W^i_{\nu}
-   \partial_{\nu}W^i_{\mu},
\quad 
\displaystyle \tilde{F}_{\mu\nu} = \frac{1}{2}
\epsilon_{\mu\nu\rho\sigma}F^{\rho\sigma}
\end{array}
\end{eqnarray}
are  introduced. The scale $\Lambda$ keeps the coupling
constant $a_i$ dimensionless.
In our
calculations $\Lambda$ are fixed by value of $M_W$ ($\sim 80$
GeV). As one can see the operators ${\cal L}_0$ and ${\cal L}_c$
are $C$-, $P$-, $CP$-invariant. $\tilde{{\cal L}}_0$ is the $P$-
and $CP$-violating operator. Then anomalous quartic boson vertices
could be defined as
\begin{eqnarray}\label{c12}
\begin{array}{l}
\displaystyle {\Gamma}_{4\,a_0}^{\mu\nu\alpha\beta}(k_1,k_2) = \frac{1}{8\Lambda^2}\left(4a_0g^{\alpha\beta}
((k_1k_2)g^{\mu\nu}-k_1^{\nu}k_2^{\mu})\right), \\  \\
\displaystyle {\Gamma}_{4\,a_c}^{\mu\nu\alpha\beta}(k_1,k_2) = \frac{1}{8\Lambda^2}\left(a_c((k_1^{\alpha}k_2^{\beta}
+k_1^{\beta}k_2^{\alpha})g^{\mu\nu}
+(k_1k_2)(g^{\mu\alpha}
g^{\nu\beta}+g^{\nu\alpha}g^{\mu\beta})-\right. \\    \\
\displaystyle\left. - k_1^{\nu}(k_2^{\beta}g^{\mu\alpha}+k_2^{\alpha}
g^{\mu\beta})- k_2^{\nu}(k_1^{\beta}g^{\mu\alpha}+k_1^{\alpha}
g^{\mu\beta}))\right), \\                               \\
\displaystyle {\Gamma}_{4\,\tilde{a}_0}^{\mu\nu\alpha\beta}(k_1,k_2) = \frac{1}{8\Lambda^2}\left(4\tilde{a}_0g^{\alpha\beta}k_{1\rho}k_{2\sigma}\
\epsilon^{\mu\rho\nu\sigma}\right).
\end{array}
\end{eqnarray}
The anomalous quartic boson vertices depend on the 4-momenta  $k_1$, $k_2$ of the photons involved in
the vertex.

Total cross section of $\gamma\gamma\rightarrow W^+W^-$ is calculated as
\begin{eqnarray}\label{c13}
\begin{array}{c}
\displaystyle \sigma_{\lambda_1\lambda_2\lambda_3\lambda_4} = \frac{1}{2S}\int|M_{\lambda_1\lambda_2\lambda_3\lambda_4}|^2
d\Gamma^{(2)},
\end{array}
\end{eqnarray}
where $M_{\lambda_1\lambda_2\lambda_3\lambda_4}$ have been defined
in  eq. (\ref{c1}), $d\Gamma$ is two-particle phase space of the
bosons.

Through the method of Monte-Carlo integration \cite{mc}
one may obtain numerical values of the cross section \cite{igor_gg_WW}.
For these purposes the events generator has been built.

\begin{figure}[h!]
\begin{minipage}[b]{.975\linewidth}
\centering
\includegraphics[width=\linewidth, height=3.8in, angle=0]{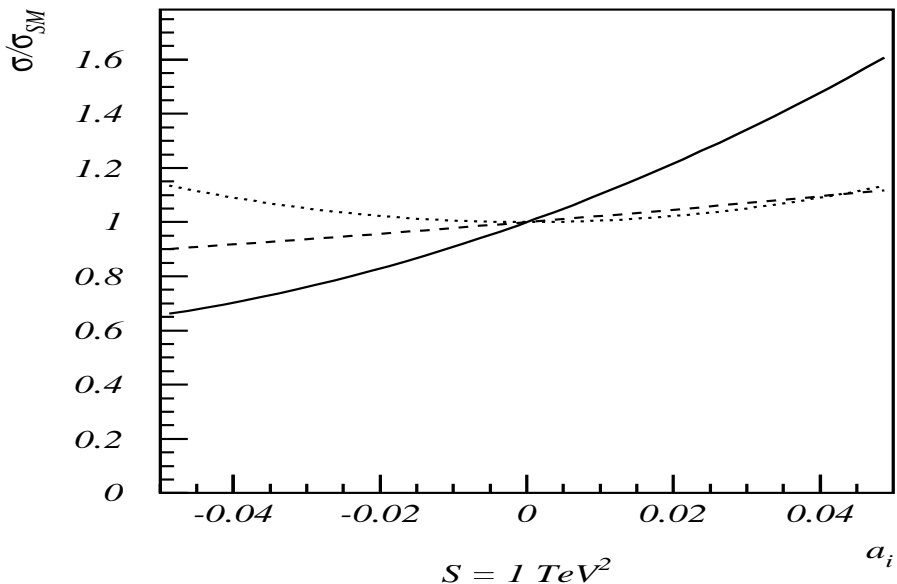}
\caption{The comparison of the dependency of the cross sections
$\sigma(W^+W^-)$. Solid line presents $a_0$-, dashed line --
$a_c$-, dotted line -- $\tilde{a}_0$-dependence} \label{f7}
\end{minipage} \hfill
\end{figure}

Fig. \ref{f7} shows the dependence of three total cross sections
$\sigma(W^+W^-)$ on anomalous parameters. Graphs presented on the
picture allow to value contributions of anomalous parts to the
cross section. The fact that the minima of the curves are close to
the SM point $a_i=0$ means that the interference between the
anomalous and the standard part of the  matrix element is very
small. Though the region of $a_i$ is small, it's about $0.05$, the
cross section with anomalous constants may reach the values of
$1.6\sigma$. That stands for the evidence of great sensitivity of
the cross section $\sigma(W^+W^-)$ to anomalous couplings. Taking
into account a luminosity ${\cal L}$ of photons about
$100fb^{-1}/year$, energy $\sqrt{s}\sim 1\, TeV$ that corresponds
to TESLA \cite{tdr}, statistical error will be equal to $0.05\%$.
\begin{figure}[h!]
\begin{minipage}[b]{.475\linewidth}
\centering
\includegraphics[width=\linewidth, height=3.8in, angle=0]{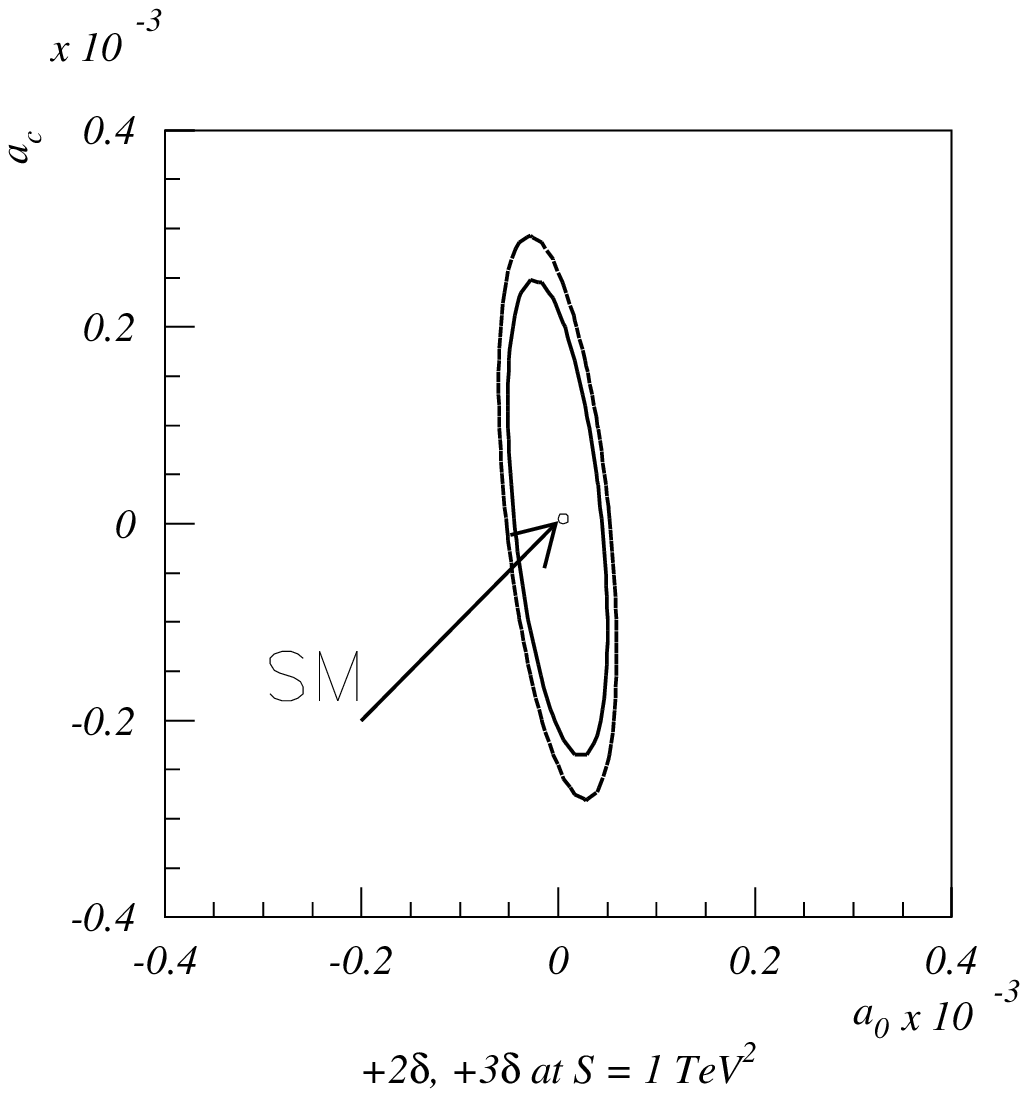}
\caption{Contour plots on ($a_0$, $a_c$) for $+2\delta$,
$+3\delta$ deviations of $\sigma(W^+W^-)$} \label{f8}
\end{minipage}
\begin{minipage}[b]{.475\linewidth} \centering
\includegraphics[width=\linewidth, height=3.8in, angle=0]{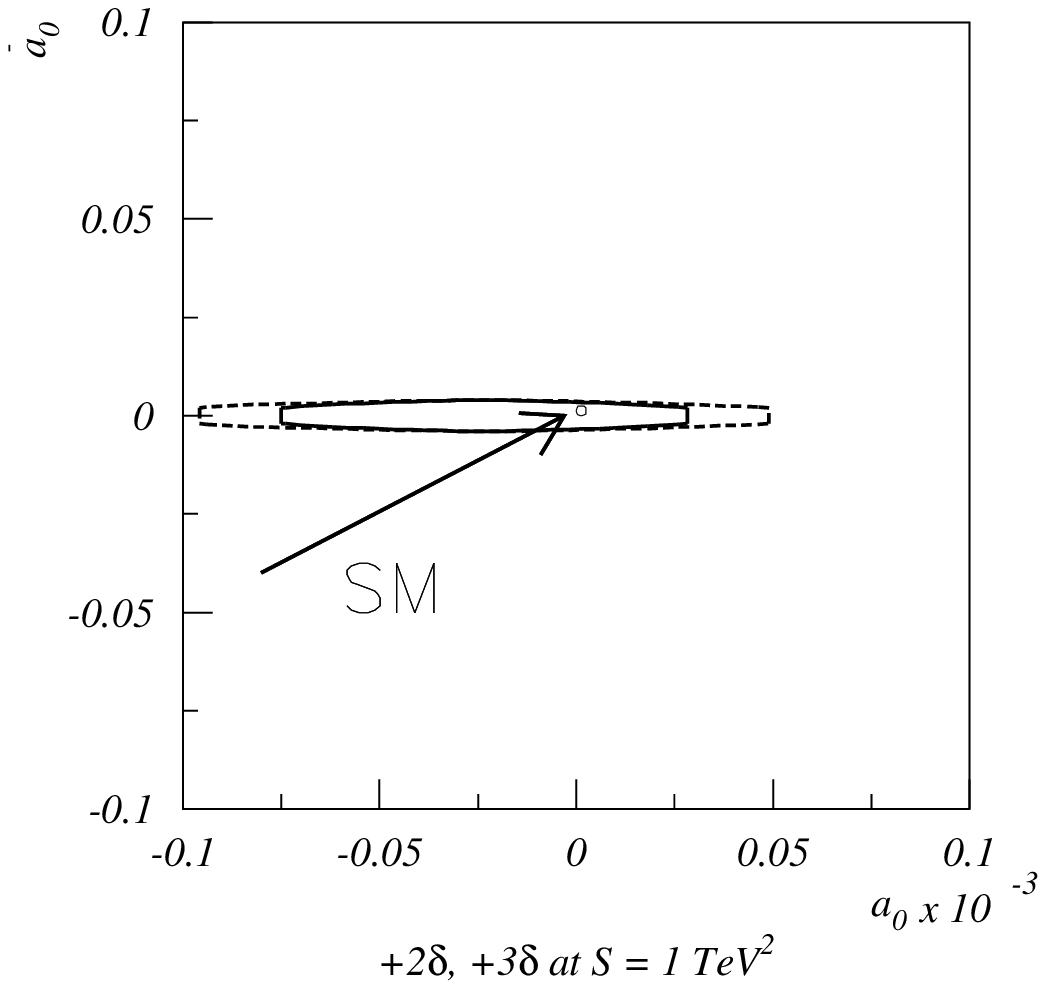}
\caption{Contour plots on ($a_0$, $\tilde{a}_0$) for $+2\delta$,
$+3\delta$ deviations of $\sigma(W^+W^-)$ } \label{f9}
\end{minipage}\hfill
\end{figure}
Therefore we can estimate bounds imposed on anomalous couplings to
compare anomalous effect with statistical error. Contour plots
on two non-zero anomalous constant, figs. \ref{f8}--\ref{f100},
clearly demonstrate these bounds.
\begin{figure}[h!]
\begin{minipage}[b]{.475\linewidth}
\centering
\includegraphics[width=\linewidth, height=3.8in, angle=0]{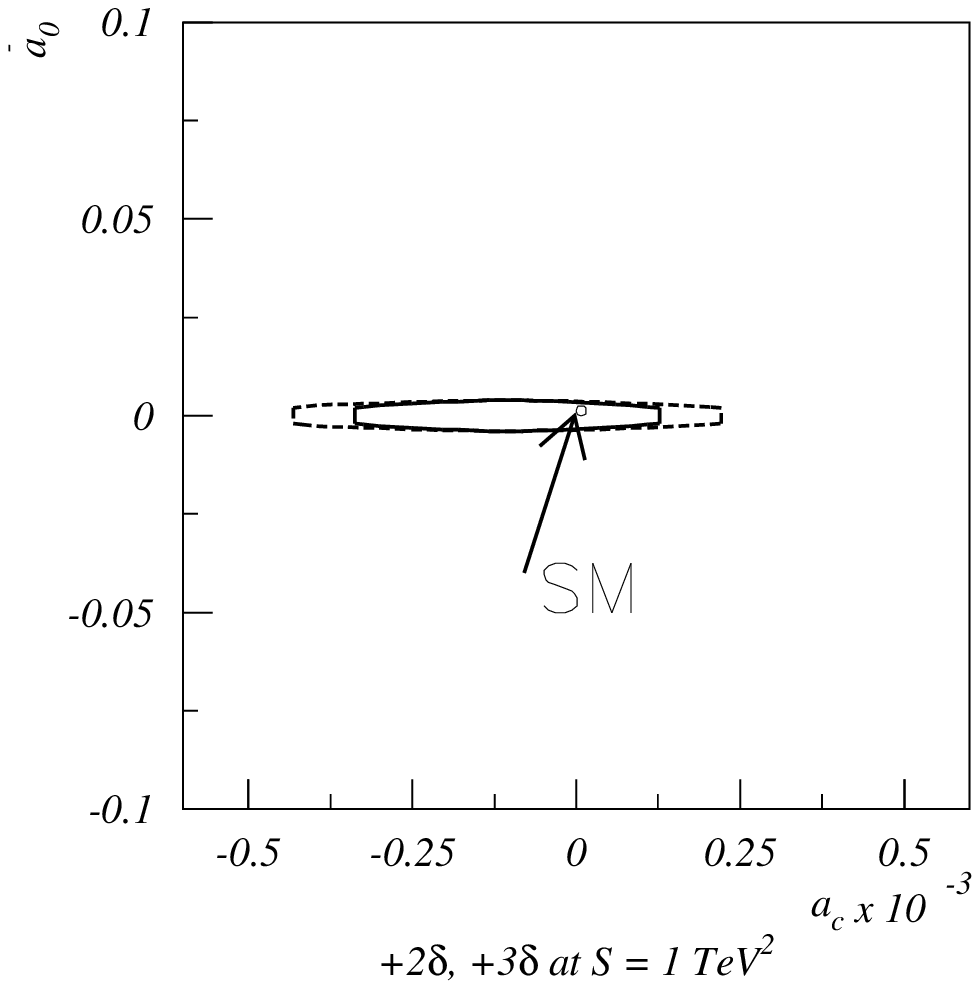}
\caption{ Contour plots on ($a_c$,$\tilde{a}_0$) for $+2\delta$,
$+3\delta$ deviations of $\sigma(W^+W^-)$} \label{f10}
\end{minipage}
\begin{minipage}[b]{.475\linewidth}
\centering
\includegraphics[width=\linewidth, height=3.8in, angle=0]{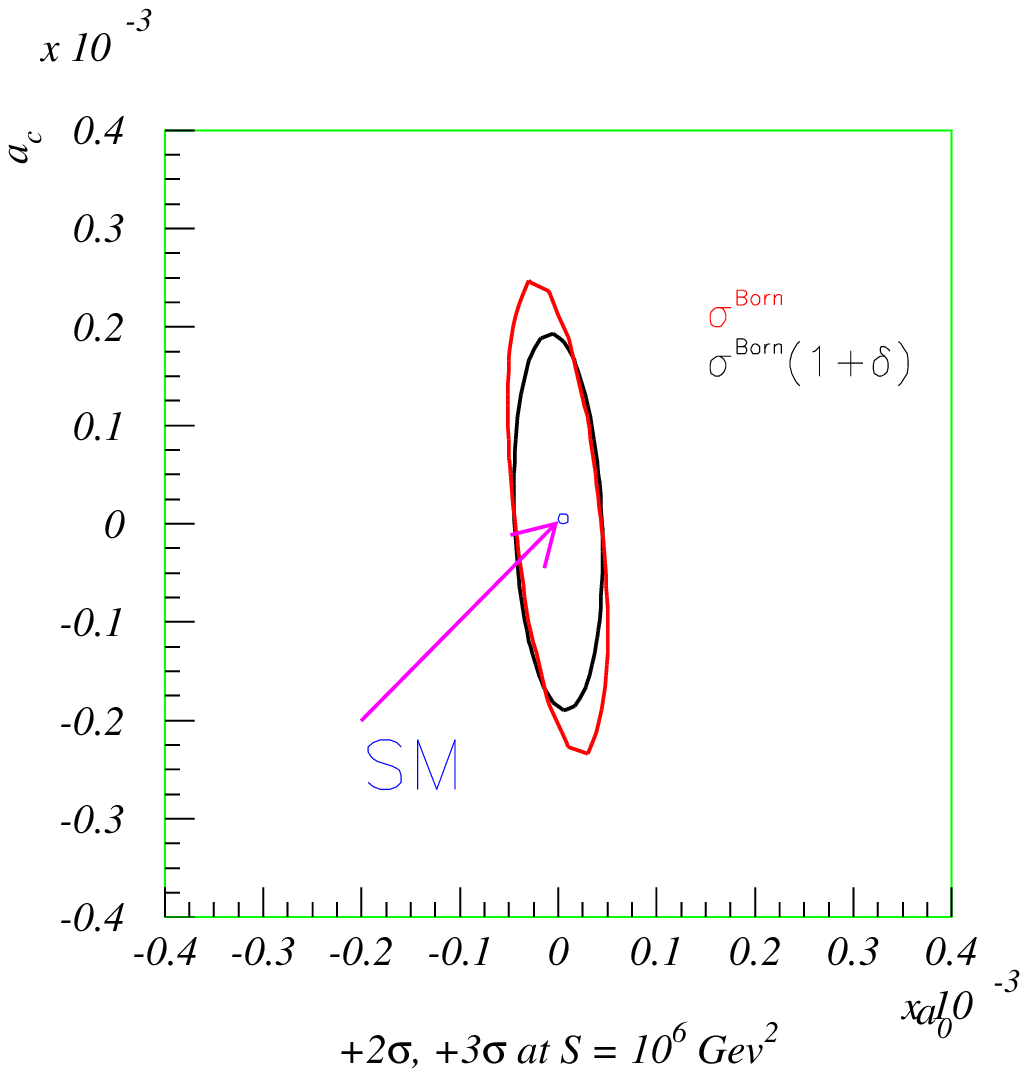}
\caption{ Contour plots on ($a_c$,$\tilde{a}_0$) for $+2\delta$
deviations of $\sigma(W^+W^-)$} \label{f100}
\end{minipage}
\end{figure}

\subsection{${\cal O}(\alpha)$ correction to anomalous constants}

Obviously   a precision  analysis of $\gamma\gamma\rightarrow WW$ at the future
high energy $\gamma\gamma$-colliders is impossible without calculation of whole set the
first-order ${\cal O}(\alpha)$ radiative corrections,
including real photon emission as well as a set of one-loop diagrams
of $\gamma\gamma\to W^+ W^-$ that are presented in fig.1 and fig.2 of ref. \cite{igor_anomal}.
Inclusive cross section of the $W$ pair production in the $\gamma\gamma$ collisions to the third order in $\alpha$ is
given by the sum of Born cross section, interference term between the Born and one-loop amplitudes and cross section of
the $WW\gamma$ production.
In case of energy of $\gamma\gamma$-interaction
exceeds threshold of three boson production
these process ($\gamma\gamma\to W^+ W^- Z$)
must be considered as radiative correction:
\begin{eqnarray}\label{c15}
\begin{array}{c}
\displaystyle
d\sigma(\gamma\gamma\rightarrow W^+W^-) = d\sigma^{Born}(\gamma\gamma\rightarrow W^+W^-) +
\frac{1}{S}Re(M^{Born}M^{1-loop *})d\Gamma^{(2)} + \\
\displaystyle
+ d\sigma^{soft}(\gamma\gamma\rightarrow W^+W^-\gamma) + d\sigma^{hard}(\gamma\gamma\rightarrow W^+W^-\gamma) +
d\sigma^{Z}(\gamma\gamma\rightarrow W^+W^-Z).
\end{array}
\end{eqnarray}
Since one-loop and soft photon emission amplitudes are
IR-divergent and only their sum is IR-finite it is convenient to
consider soft and hard photon emissions separately.
$d\sigma^{soft}$ can be presented by factorizable expression
\begin{eqnarray}\label{c16}
\begin{array}{c}
\displaystyle
d\sigma^{soft}(\gamma\gamma\rightarrow W^+W^-\gamma) = d\sigma^{Born}(\gamma\gamma\rightarrow W^+W^-)R^{soft} (\omega),
\end{array}
\end{eqnarray}
where $\omega$ is soft photon energy cutoff, $\beta = \sqrt{1-4m_W^2/s}$.
The differential cross section of hard photon emission is given by
\begin{eqnarray}\label{c18}
\begin{array}{c}
\displaystyle
d\sigma^{hard}(\gamma\gamma\rightarrow W^+W^-\gamma) = d\sigma(\gamma\gamma\rightarrow W^+W^-\gamma) - d\sigma^{soft}(\gamma\gamma\rightarrow W^+W^-\gamma)
\end{array}
\end{eqnarray}
and can not be factorized.
$d \sigma^{soft}$ and $d \sigma^{hard}$ present two contributions independent from infrared divergence
as well as from soft photon cutoff.

One-loop amplitude $M^{1-loop}$ were built due to usage of SCA
(Algebra of Symbolic Calculations) programs ($MATHEMATICA,\,
REDUCE...$) and transformations of scalar and tensor integrals to
one-, two-, ..., six-point integrals. Cross section of $WWZ$
production on $\gamma\gamma$ beams are obtained through application
of the Monte-Carlo method of numerical integration and exact
covariant expressions of $\gamma\gamma\rightarrow W^+W^-Z$
amplitudes.
Cross section of $\gamma\gamma\to W^+ W^-$ including the lowest-order QED radiative correction
and three-boson production one can see in refs.
\cite{igor_gg_WW, igor_anomal, igor_quartic}.

Contour plot on $a_0$ and $a_c$ of $\sigma(WW)$ with the lowest
order QED correction are presented in figs. \ref{f8}-\ref{f100}.

\section{Conclusion}

We have analyzed the possible usage of $\gamma \gamma \to f
\bar{f} \gamma$ reaction for the luminosity measurement at
$J\!=\!0$ beams on linear photon collider. The achievable
precision of the luminosity measuring is considered and
calculated. The optimal conditions for that measurement are found
(for the high magnitude of $J\!=\!0$ cross section and small
$J\!=\!2$ background). The first-order QED correction to
$\gamma\gamma\to l \bar{l}$ cross section is calculated and analyzed
at $J\!=\!2$-beams.

The considered process gives the excellent opportunity for luminosity measurements with substantial accuracy.

The cross section of process of four charged lepton production are
obtained in frame of method
of helicity amplitudes as well as method of
covariant calculations.
Comparative analysis are performed.
Monte-Carlo generator was
developed to calculate
differential and total cross section
of four lepton production.

The investigation of the sensitivity of process of
$\gamma\gamma\to W^+ W^-$ and $\gamma\gamma\to W^+ W^- Z$ to
genuine anomalous quartic couplings $a_0$, $a_c$ and $\tilde{a}_0$
was performed at center-of-mass energy $\sqrt{s}=1\, TeV$. It was
discovered that two-boson production has great sensitivity to
anomalous constants
$a_c$ and $a_0$ but process $\gamma\gamma\to W^+ W^- Z$ is more
suitable for study of $\tilde{a}_0$.

The fact that the minimum of the curves are close to the SM point $a_i=0$ demonstrates
the small value of the
anomalous and the standard part interference.
The first-order radiative correction to
cross section $\sigma (\gamma\gamma\to W^+ W^-)$ has significant
magnitude and its calculation increases the precision
of the $a_0$ and $a_c$ measurement.

\end{document}